\documentclass{aa}  
\usepackage{graphicx}
\usepackage{relsize}
\usepackage{txfonts}
\usepackage{xcolor}
\usepackage{fancyvrb}
\usepackage{multirow}   
\usepackage{placeins}
\usepackage{stfloats}

\definecolor{bblue}{HTML}{002099} 
\definecolor{rred}{HTML}{F22616}  
\definecolor{ggren}{HTML}{00693E} 

\usepackage[breaklinks,colorlinks,citecolor=bblue,linkcolor=bblue,urlcolor=bblue]{hyperref}
\usepackage[capitalise]{cleveref}

\usepackage[all]{hypcap} 

\defcitealias{Ogle2024}{O24}
\defcitealias{GarciaBurrillo2009A&A...496...85G}{GB09}

\begin{document} 
   \title{Shocked, heated, and now resolved:\\ H$_2$ excitation in the low-luminosity AGN at M58 core with JWST}
        \author{I.\,E.\,L\'opez\inst{\ref{ins:oas}}\thanks{e-mail: ivan.lopez@inaf.it}
        \and
        E.\,Bertola\inst{\ref{ins:arcetri}}
        \and
        V.\,Reynaldi\inst{\ref{ins:ialp},\ref{ins:obser}}
        \and
        P.\,Ogle\inst{\ref{ins:stsci}}
        \and
        R.\,D.\,Baldi\inst{\ref{ins:ira}}
        \and
        M.\,Brusa\inst{\ref{ins:difa},\ref{ins:oas}}
        \and
        S.~Garc\'{\i}a-Burillo\inst{\ref{ins:oan}}
        \and
        B.\,Sebastian\inst{\ref{ins:stsci}}
        \and
        M.\,V.\,Zanchettin\inst{\ref{ins:arcetri}}
        \and
        G.\,Cresci\inst{\ref{ins:arcetri}}
        \and
        J.\,A.\,Fern\'andez-Ontiveros\inst{\ref{ins:teruel}}
        \and
        A.\,Marconi\inst{\ref{ins:firenze},\ref{ins:arcetri}}
        \and
        R.\,M.\,Rich\inst{\ref{ins:cali}}
        \and
        T.\,M.\,Rodriguez\inst{\ref{ins:cologne}}
        }

   \institute{
            INAF - Osservatorio di Astrofisica e Scienza dello Spazio di Bologna, via Gobetti 93/3, 40129, Bologna, Italy\label{ins:oas}
            \and
            INAF - Osservatorio Astrofisco di Arcetri, largo E. Fermi 5, 50127, Firenze, Italy\label{ins:arcetri}
            \and
            Instituto de Astrofísica de la Plata (CONICET—UNLP), Paseo del Bosque, 1900, La Plata, Argentina\label{ins:ialp}  
            \and
            Facultad de Ciencias Astronómicas y Geofísicas, Universidad Nacional de La Plata, Paseo del Bosque, 1900, La Plata, Argentina\label{ins:obser} 
            \and
            Space Telescope Science Institute, 3700 San Martin Drive, Baltimore, MD, 21218, USA\label{ins:stsci}    
            \and
            INAF - Istituto di Radioastronomia, via Gobetti 101, 40129, Bologna, Italy\label{ins:ira}
            \and
            Dipartimento di Fisica e Astronomia "Augusto Righi", Università di Bologna, via Gobetti 93/2, 40129, Bologna, Italy \label{ins:difa}
            \and
            Observatorio Astron\'omico Nacional (OAN-IGN)-Observatorio de Madrid, Alfonso XII 3, 28014, Madrid, Spain\label{ins:oan}
            \and 
            Centro de Estudios de Física del Cosmos de Aragón (CEFCA), Plaza San Juan 1, 44001, Teruel, Spain \label{ins:teruel}
            \and
            Dipartimento di Fisica e Astronomia, Universitá di Firenze, via G. Sansone 1, 50019, Firenze, Italy\label{ins:firenze}    
            \and
            Department of Physics and Astronomy, University of California, Los Angeles, CA  90095-1547 USA\label{ins:cali}  
            \and
            I. Physikalisches Institut, Universit\"{a}t zu K\"{o}ln, Z\"{u}lpicher Stra{\ss}e 77, 50937, Cologne, Germany\label{ins:cologne}
            }


\abstract{
We present JWST NIRSpec and MIRI MRS observations of the central kiloparsec of M58 (NGC\,4579), a nearby galaxy hosting a low-luminosity AGN (LLAGN; $L_\mathrm{bol}$\,$\sim$\,$10^{42}$\,erg\,s$^{-1}$) with a low-power jet. These data provide an unprecedented view of the warm molecular gas phase and reveal clear signatures of feedback. We detect 44 H$_2$ lines, including bright pure rotational lines (S(1)–S(18)) and rovibrational lines up to $\nu$\,=\,$2$, probing a wide range of excitation conditions. Excitation diagrams show that rotational lines follow a power-law temperature distribution with an exponential cutoff, consistent with heating by low-velocity shocks. H$_2$ rovibrational lines deviate from thermal models primarily because of sub-thermal excitation at low density. Additionally, there may be a 10\% contribution powered by AGN X-ray photons in the nucleus. The dust lanes associated with the spiral inflow appear dynamically undisturbed but show signs of shock heating, while the inner $\sim$200 pc exhibits turbulent kinematics produced by outflowing molecular gas. These results reveal the subtle yet measurable impact of LLAGN feedback on the interstellar medium, demonstrating that even weak, vertically oriented jets and low radiative accretion rates can perturb molecular gas and regulate nuclear reservoirs. This study highlights JWST’s transformative ability to uncover hidden modes of AGN feedback.}


\titlerunning{H$_2$ excitation in the low-luminosity AGN at M58 core}
\maketitle

\section{Introduction}
\label{sec:1}


Molecular hydrogen (H$_2$), the most abundant molecule in the Universe, is paradoxically difficult to observe directly. As a homonuclear diatomic molecule, it lacks a permanent dipole moment, and its quadrupolar rotational and vibrational transitions are intrinsically weak. These lines arise only when the gas is warm ($\sim$\,200--5000\,K) but not yet dissociated. Diagnostic ratios of pure rotational and rovibrational transitions offer sensitive probes of excitation mechanisms, offering insights into the interplay between the interstellar medium (ISM), star formation (SF), and black hole activity \citep[e.g.,][]{Mouri1994ApJ...427..777M,Roussel2007ApJ...669..959R}.

In active galactic nuclei (AGN), black hole accretion drives complex feedback that affects the ISM of the host galaxy \citep[e.g.,][]{HarrisonRamos2024Galax..12...17H}. In high-accretion systems, luminous AGN emit intense radiation producing ionization cones and extended outflows \citep[e.g.,][]{Cicone2014A&A...562A..21C,Brusa2018A&A...612A..29B}. Conversely, low-accretion AGN can show collimated jets or slower winds that mechanically interact with the ISM \citep[e.g.,][]{Morganti2005A&A...444L...9M,HeckmanBest14}. These radiative and kinetic feedback modes are critical components of cosmological simulations, which require AGN feedback to regulate SF and reproduce black hole–galaxy scaling relations \citep[for a review, see][]{Somerville2015ARA&A..53...51S}.

Outflows driven by AGN are observed to be multiphase from ionized to cold molecular components. Yet, whether this feedback quenches or triggers SF remains debated. Some AGN hosts show suppressed SF or gas depletion \citep[e.g.,][]{Lammers2023ApJ...953...26L,Bertola2024A&A...691A.178B,Garcia2024A&A...689A.347G}, while others lie on or above the main sequence \citep[e.g.,][]{Mullaney2012ApJ...753L..30M,Jarvis2020MNRAS.498.1560J}. Further complicating the picture, radiative and mechanical feedback can act simultaneously \citep[e.g.,][]{Harrison2015ApJ...800...45H}, and their impact can vary spatially and temporally within the same galaxy \citep[e.g.,][]{Bessiere2022MNRAS.512L..54B}. 

\begin{figure*}
    \centering
    \includegraphics[width=1\linewidth]{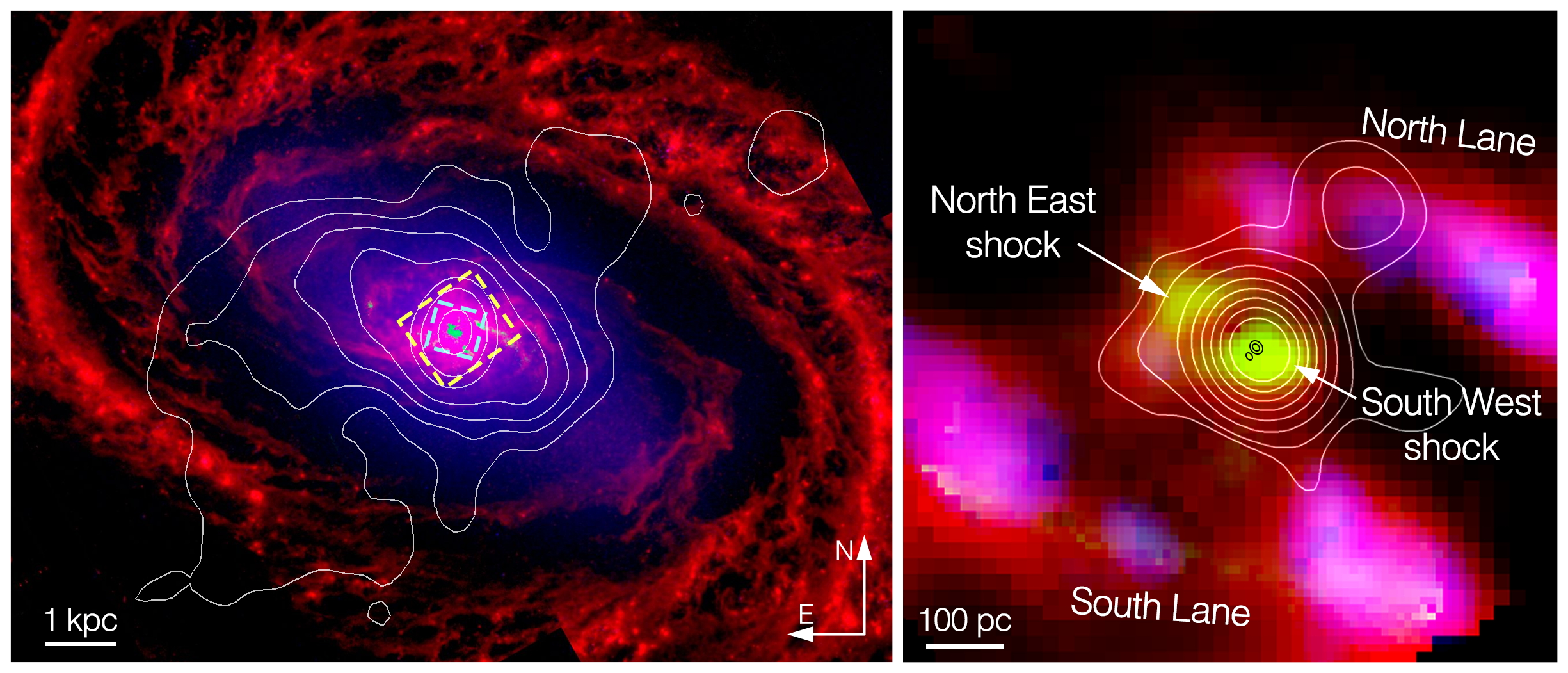}
    \caption{\textit{Left:} Composite image showing F200W (blue; stellar continuum), continuum-subtracted H$_2$ 1–0 S(1) from F212N (green), and F770W (red; tracing PAH 7.7 $\mu$m emission). The LOFAR 144 MHz radio contours are overlaid. The NIRSpec and MIRI Channel 4 fields of view are shown with cyan and yellow dashed lines, respectively. \textit{Right:} Zoom-in highlighting warm molecular gas traced by H$_2$ S(9) (green), S(1) (red), and CO 2-1 (blue). Overlaid contours are C band from VLA (white) and MERLIN (black). Key regions are labeled, including the northern and southern dust lanes as well as shocked regions to the northeast and southwest. Physical scale bars are shown in both panels.}
    \label{fig:fov}
\end{figure*}

When AGN inject mechanical energy into the ISM, they dominate the excitation of molecular gas. H$_2$ lines thereby become the most luminous mid-IR features, as seen in Stephan’s Quintet \citep{Appleton2006ApJ...639L..51A,Appleton2023ApJ...951..104A}. In radio-loud AGN, Spitzer revealed a population of molecular hydrogen emission galaxies (MOHEGs), where warm H$_2$ emission is bright, extended, and decoupled from SF \citep{2007ApJ...668..699O,2010ApJ...724.1193O}. These galaxies show extreme H$_2$ to polycyclic aromatic hydrocarbon (PAH) ratios indicating shock excitation rather than UV heating. The most plausible driver is mechanical heating from jets, possibly aided by cosmic rays, while X-ray dissociation regions (XDRs) fall short of explaining the observed emission \citep{2010ApJ...724.1193O}.

Low-luminosity AGN (LLAGN; $L_X$\,$<$\,$10^{42}$\,erg\,s$^{-1}$), including low-ionization nuclear emission-line regions (LINERs) and weak Seyferts, are typically powered by radiatively inefficient accretion flows (RIAFs), such as advection-dominated accretion flows (ADAFs) or truncated thin disks \citep[for a review, see][]{YuanNarayan2014ARA&A..52..529Y}. These flows emit little UV radiation but can launch powerful megaparsec-scale jets that affect the intergalactic medium, down to low-power radio jets ($P_\textrm{jet}\sim10^{42}$--$10^{44}$\,erg\,s$^{-1}$) confined to galactic scales \citep[e.g.,][]{Baldi18LemmingsI,mingo19,pierce20,Baldi21LemmingsII,webster21}. These jets interact with the ISM, driving shocks that excite the gas \citep[e.g.,][]{Guillard2012ApJ...747...95G}. Low-luminosity AGN with low-power radio jets are valuable laboratories for studying gentle, long-lived feedback \citep[e.g.,][]{Sabater19,khar23,Ulivi2024A&A...685A.122U}. Though they lack the dramatic outflows of quasars and powerful radio galaxies, LLAGN dominate the local AGN population and account for most of a black hole’s lifetime \citep{Ho08,Novak2011ApJ...737...26N,Burke2025ApJ...978...77B}. Often dismissed as negligible \citep{Shin2019ApJ...881..147S}, growing evidence suggests that they can drive turbulence and inflate radio bubbles, shocking the multiphase gas on kiloparsec scales \citep[e.g.,][]{Alatalo2011ApJ...735...88A,Mezcua2014ApJ...787...62M,Goold2024ApJ...966..204G}.

Messier 58 (NGC\,4579) illustrates the impact of low-power jet feedback on a nearby galaxy. Located in the Virgo Cluster at 21\,Mpc, it hosts a LINER nucleus \citep{balmaverde16} powered by an ADAF \citep{Nemmen14}, which launches a radio jet with an estimated kinetic power of $\sim2\times10^{43}$\,erg\,s$^{-1}$, derived from its 1.4 GHz luminosity ($\nu S_\nu$\,$=$\,$1.4 \times 10^{39}$\,erg\,s$^{-1}$) and the relation by \citet{2010ApJ...720.1066C}. Spitzer IRS observations revealed a bright warm H$_2$ disk extending $\sim$2.6 kpc \citep[][hereafter O24]{Ogle2024}. CO observations show disrupted molecular gas lanes spiraling inward toward the nucleus \citep[][hereafter GB09]{GarciaBurrillo2009A&A...496...85G}, and Gemini near-IR imaging shows that warm H$_2$ emission is bright along these lanes. Central H$_2$ line ratios indicate thermal excitation \citep{Mazzalay2013MNRAS.428.2389M}. Jet-driven heating is consistent with simulations where a low-power jet perpendicular to the disk inflates a hot bubble into the ISM \citep{2016MNRAS.461..967M,2018MNRAS.479.5544M}, driving shocks and exciting H$_2$ as it interacts with dust lanes. The resulting cocoon extends over kiloparsec scales and coincides with suppressed SF \citepalias{Ogle2024}.

The proximity of \object{M58} makes it a prototypical system for studying LLAGN feedback, particularly in the context of MOHEGs. We were awarded a comprehensive JWST program totaling $\sim$17\,hours with NIRCam, NIRSpec, and MIRI to dissect jet–ISM interactions and construct a multiphase view in this galaxy. JWST’s unprecedented spatial and spectral resolution enables detailed probing of the influence of both jet and ADAF. This paper focus on the excitation and dynamics of warm H$_2$ emission, setting the stage for a deeper understanding of LLAGN feedback. Companion papers will address the ionized gas and PAH emission, completing the multiphase picture. 

Throughout this work, we adopt a distance of 21\,Mpc based on SN\,1989M \citep{1996ApJ...465L..83R}. All maps are shown with north up and east to the left. We use the notation $v_\mathrm{up}$–$v_\mathrm{low}$\,X($J_\mathrm{low}$) for H$_2$ lines, where $v$ and $J$ denote the vibrational and rotational quantum numbers and X indicates the branch (S, Q, or O). Pure rotational lines are denoted simply as S($J_\mathrm{low}$).


\section{Observations and data analysis}
\label{sec:2}

M58 was observed with JWST in June 2024 as part of program 3671 (PI: I.E. L\'opez), using the full suite of instruments. JWST integral field units (IFU) observations were carried out as 2$\times$2 mosaics with NIRSpec and MIRI MRS. \Cref{fig:fov} shows an overview of the galaxy with the IFU footprints. NIRSpec observations used the G235H F170LP and G395H F290LP grating and filter pairs, covering 1.66–5.27\,$\mu$m, with each mosaic including two sets of four-point medium cycling dithers, totaling 31 minutes of integration per grating. MIRI MRS data observations cover 5–28\,$\mu$m and used a four-point dither pattern in slow readout mode, with 1.6 hours total integration, including dedicated background frames. Imaging observations with NIRCam and the MIRI Imager used a combination of narrow-, medium-, and broad-band filters (F200W, F212N, F300M, F335M, F770W, F1000W, F1130W, and F1280W) to trace stellar, H$_2$, and dust emission, providing a broader morphological context and supporting astrometric correction of the IFU data.

\begin{figure*}
    \centering
    \includegraphics[width=0.95\linewidth]{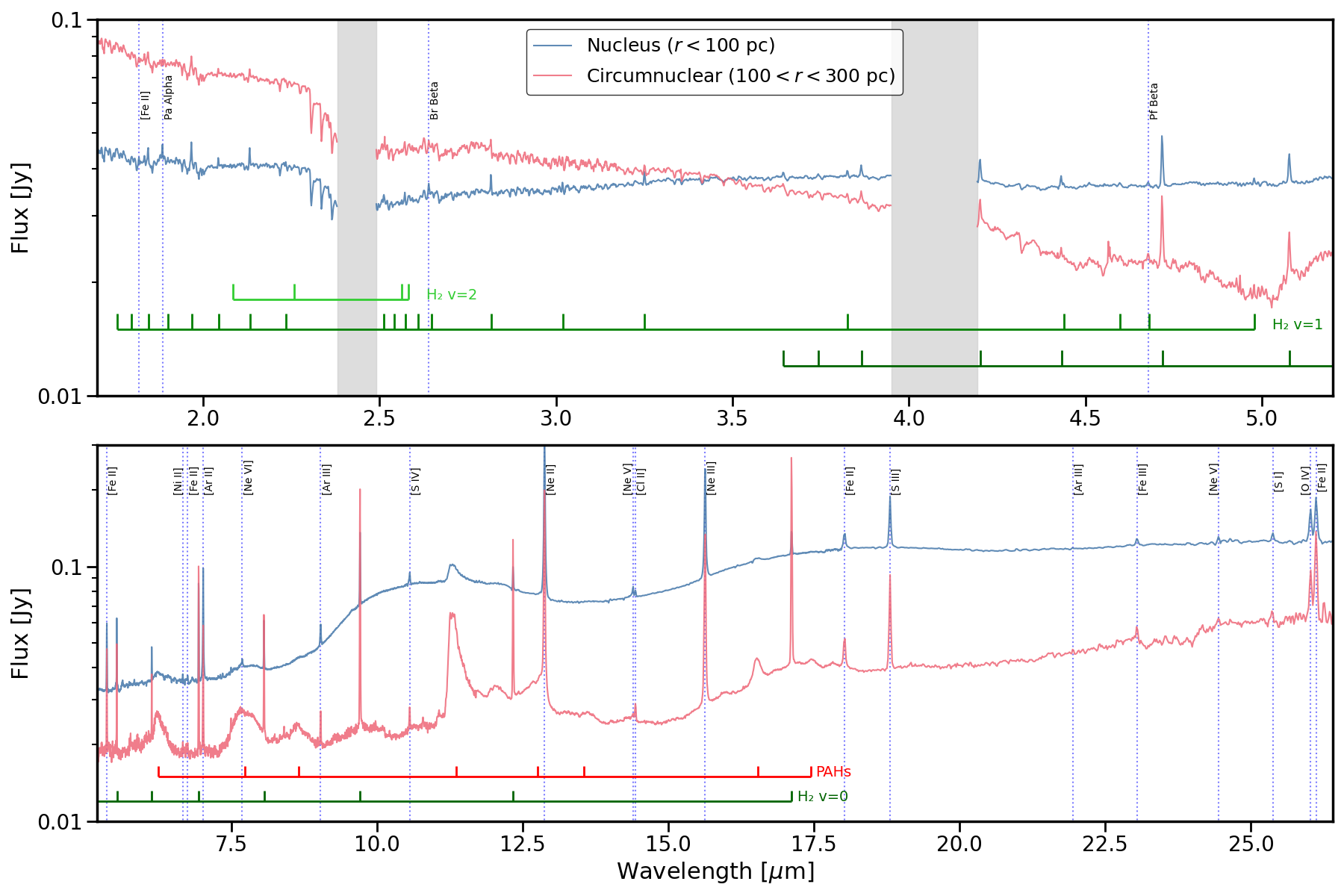}
    \caption{JWST spectra for the nucleus (blue) and for a circumnuclear region (red). Ionic emission lines are indicated by vertical blue lines. The PAH features and H$_2$ lines are marked along the bottom of each panel. The circumnuclear region exhibits strong PAH emission, pure rotational H$_2$, and low-excitation ionized lines. In contrast, the nuclear spectrum shows prominent high-ionization lines, relatively stronger H$_2$ rovibrational lines, and two silicate features at 10 $\mu$m and 17 $\mu$m. Shaded-gray regions indicate gaps in the NIRSpec instrument.}
    \label{fig:spectra}
\end{figure*}

Reduction of the datasets followed the JWST Science Calibration Pipeline, with instrument-specific customizations. Full details on data reduction and correction for noise sampling features and fringing are provided in \Cref{appendix:data_red}. Spectral fitting was performed via multicomponent spaxel modeling, detailed in \Cref{appendix:fitting}. The velocity center adopted throughout the analysis corresponds to the redshift of the source ($v_{\mathrm{hel}} = 1518$\,km\,s$^{-1}$).

We used archival MERLIN and VLA C-band (Project ID 20A-272, PI: Ogle) observations to trace the radio structure. The MERLIN map reveals an unresolved 5\,GHz core with a flux density of 40$\pm$2 mJy beam$^{-1}$ and a fainter component (2.4$\pm$0.5 mJy beam$^{-1}$). VLA B-configuration imaging shows a jet-like extension on arc second scales, consistent with previous VLA maps \citep{2001ApJS..133...77H}. Together, the radio data indicate a jet inclined by $\sim$56° from the plane of the sky. Details of the data reduction are provided in \cref{appendix:data_red}. Recent LOFAR 144\,MHz observations reveal a large-scale swirling feature, in a counterclockwise direction and opposite to that of the spiral arms, corresponding to the extended jet structure \citep{Edler2023A&A...676A..24E}. 

Ultraviolet imaging was obtained with HST, using the ACS/HRC F250W filter (Program ID 9454; PI: Maoz). We combined two exposures and applied an astrometric correction of 0\farcs2 to align the UV core with the MERLIN radio continuum, as the astrometry was limited primarily by guide star uncertainties. The UV morphology reveals a ring-shaped structure interpreted as an ultracompact nuclear ring of young stars with a radius of $\sim$150\,pc \citep{2008A&A...478..403C}. We also used Spitzer IRS maps of the H$_2$ S(3) line from \citetalias{Ogle2024} and CO(2–1) observations from the Plateau de Bure Interferometer (PdBI) presented by \citetalias{GarciaBurrillo2009A&A...496...85G}.

\section{Results}
\label{sec:3}

\begin{figure*}
    \centering
    \includegraphics[width=1\linewidth]{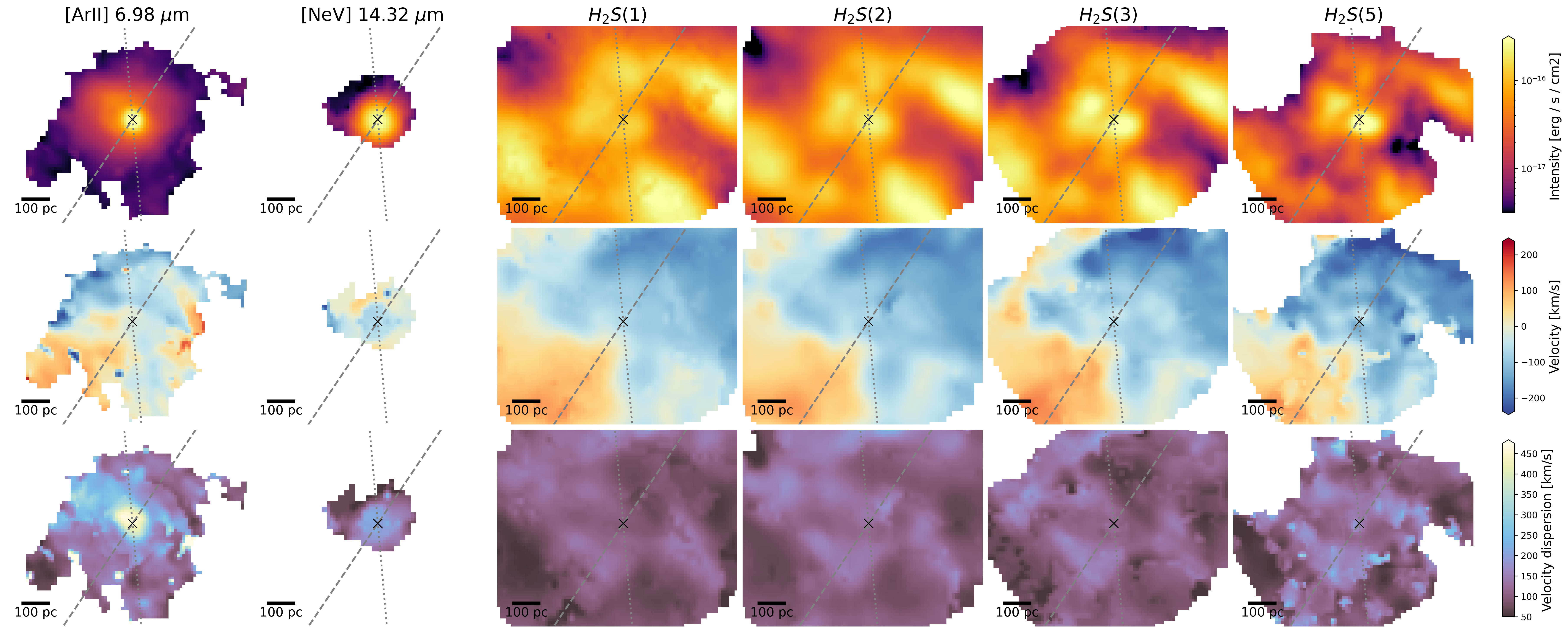}\\
    \includegraphics[width=1\linewidth]{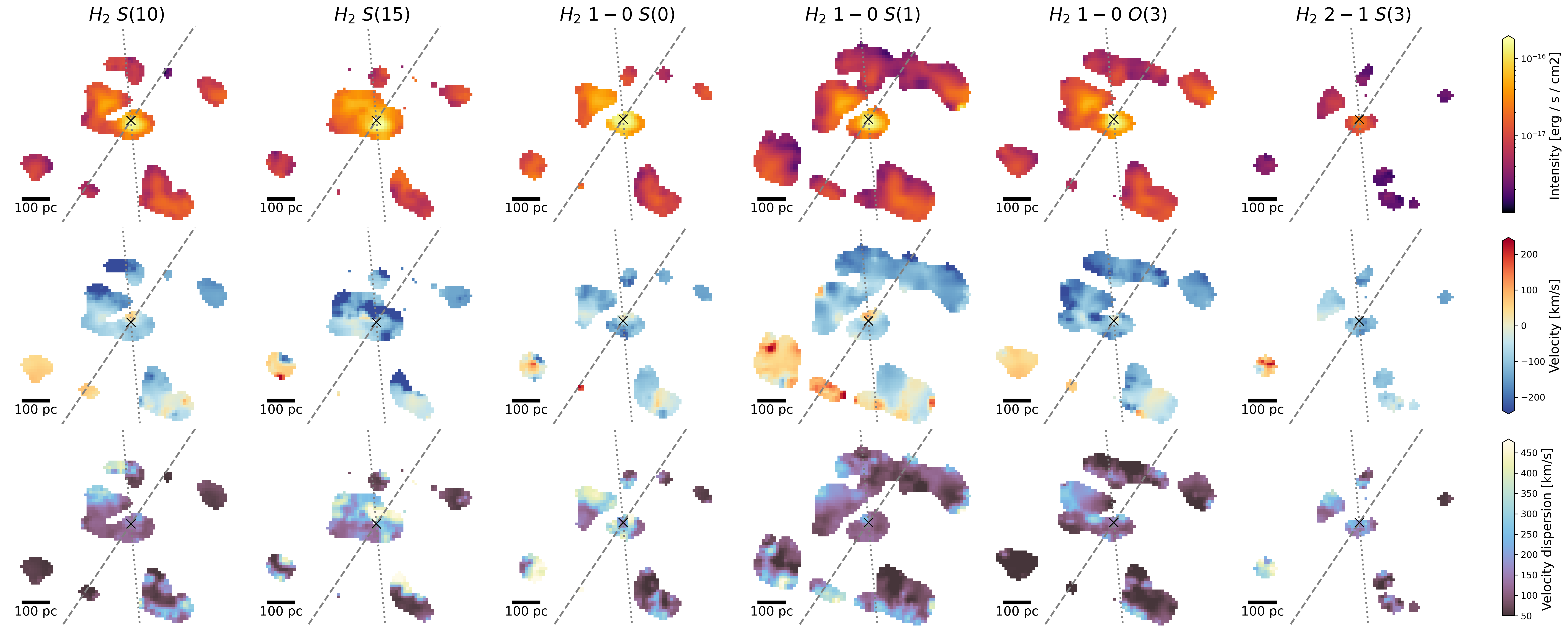}\\
    \caption{Moment maps showing H$_2$, low-ionization [Ar\,\textsc{ii}], and high-ionization lines [Ne\,\textsc{v}]. Only pixels with detections above 5$\sigma$ are shown. The dashed and dotted gray lines indicate the jet axis and the galaxy’s major axis, respectively. A black x marker indicates the AGN radio core.}
    \label{fig:mom_miri}
\end{figure*}

JWST probes the inner kiloparsecs of M58 with unprecedented detail (\cref{fig:fov}). The PAH~7.7\,$\mu$m emission traces the dusty spiral arms and the inner dust disk. NIRCam F200W imaging maps the old stellar population as a smooth continuum aligned with a $\sim$\,8 kpc bar and decoupled from the inner dust disk. The F212N narrowband show the H$_2$\,1–0\,S(1) rovibrational emission confined to the inner 2.6 kpc and consistent with \citetalias{Ogle2024}. The IFU data cover the central 0.7–1.2 kpc and trace the molecular phase of the ISM around the AGN. Two main dust lanes to the north and south dominate the H$_2$ emission, converging toward the nucleus, in agreement with absorption seen in HST optical images and CO morphology \citepalias{Ogle2024}. Closer to the AGN, two regions show enhanced H$_2$ emission: one in the northeast (NE) at $\sim$100 pc and another in the southwest (SW), partially overlapping the nucleus.

The nuclear spectrum (\cref{fig:spectra}), extracted within a 1\arcsec\ aperture ($\approx$\,100\,pc), is dominated by a strong mid-IR continuum and exhibits two prominent silicate emission features (9-13\,$\mu$m and 15-19\,$\mu$m). Although the bright continuum partially masks the PAH bands, the 11.3$\mu$m feature is clearly detected. Strong lines from ionized species ([Ne\,\textsc{v}], [Ar\,\textsc{ii}], [Fe\,\textsc{ii}]) and numerous H$_2$ transitions (pure rotational and high-$\nu$ rovibrational lines) indicate a mix of photoionization and mechanical heating. Circumnuclear spectra, extracted from a ring between 1\arcsec\ and 3\arcsec, show that both H$_2$ rotational lines and low-ionization lines ([Ar\,\textsc{ii}], [Fe\,\textsc{ii}], [Ne\,\textsc{ii}], and [S\,\textsc{iii}]) are strong. The CO absorption bands are also detected across the field. A list of all detected H$_2$ lines for the different regions is given in \Cref{table:h2}.

Moment maps (\cref{fig:mom_miri}) show that H$_2$ emission follows the dust lanes, except for very high-excitation transitions (e.g., $\nu$\,$=$\,$2$). The low-ionization line [Ar\textsc{ii}] extends within the inner 200\,pc and peaks at the AGN. The high-ionization line [Ne\textsc{v}] is confined to $\sim$100 pc and displays distinct kinematics. The enhanced mid-$J$ H$_2$ emission (e.g., S(5)) relative to lower-$J$ lines (e.g., S(1)) is consistent with \citetalias{Ogle2024}. However, the presence of higher vibrational levels within $\lesssim$\,100\,pc of the AGN suggests that additional excitation sources may contribute (see \cref{sec:Discussion}). The velocity fields of low-ionization lines and H$_2$ appear broadly consistent, with a $\pm$200\,km\,s$^{-1}$ difference between the northern and southern lanes. While this may suggest rotation, caution is advised due to noncircular motions driven by the stellar bar \citepalias{GarciaBurrillo2009A&A...496...85G}. Nevertheless, the lanes show low velocity dispersion as expected for rotation, whereas the nucleus reaches several hundred km\,s$^{-1}$. Ionized and high-excitation H$_2$ lines show elevated dispersions in the nucleus and NE.

\begin{figure*}
    \centering
    \includegraphics[width=0.95\linewidth]{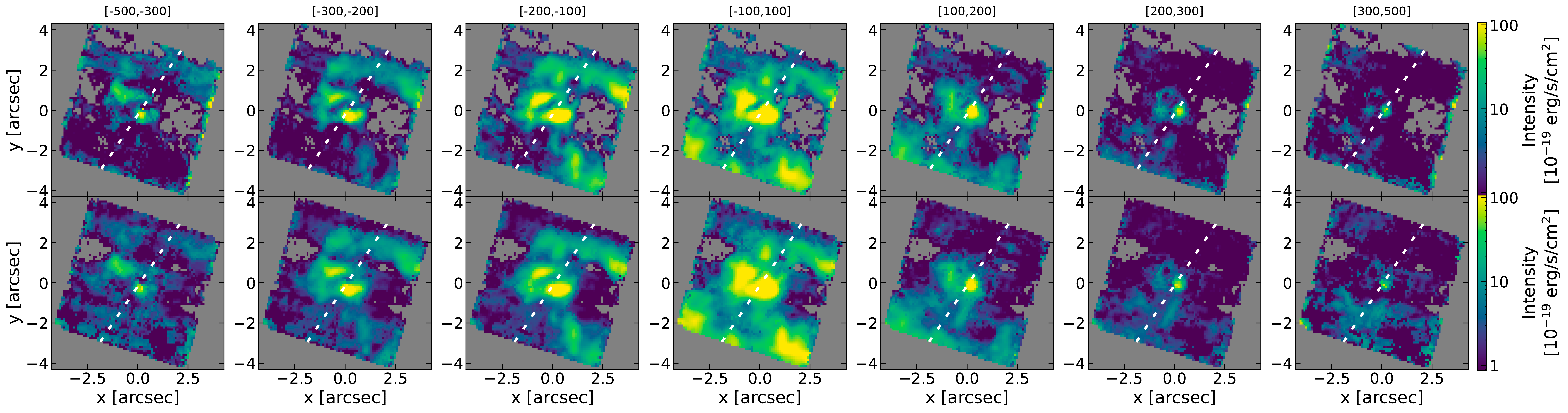}
    \caption{NIRSpec channel maps, illustrating both rotational and rovibrational kinematics. \textit{Top row:} H$_2$\,1–0 S(1); \textit{Bottom row:} H$_2$\,S(9). 
    Each column corresponds to a velocity bin (from –500 to +500 km s$^{-1}$). A dashed white line represents the jet axis. Both transitions trace the dust lanes between –300 to +200 km s$^{-1}$, compatible with rotation. The NE feature, strongest at –500 km s$^{-1}$ and persisting to +200 km s$^{-1}$, is identified as the forward shock front. The overall lack of alignment with the jet axis suggests the jet passes outside the disk plane.}
    \label{fig:channels}
\end{figure*}

The NIRSpec channel maps of the H$_2$\,1–0\,S(1) and H$_2$\,S(9) transitions (\cref{fig:channels}) show that the two main dust lanes (north and south) appear in a limited number of bins, with reduced velocity dispersion, consistent with a rotating disk. The northern lane spans –300 to –100 km\,s$^{-1}$, while a southern bridge connecting the nucleus to the southern lane appears between –200 and –100 km\,s$^{-1}$. The main southern lane itself spans from –100 to +200 km\,s$^{-1}$. The NE feature is strongest at –500 km\,s$^{-1}$ and persists up to +200 km\,s$^{-1}$. The AGN nucleus emits across nearly all bins, showing high velocity dispersion and indicating broad, turbulent motions. The emission shows a complex morphology: it starts as a vertical column between –500 and –300 km\,s$^{-1}$, takes on a curved arc shape across most bins, and eventually splits into two lobes at its outer edges above +200 km\,s$^{-1}$.

We do not find any strong alignment of velocity features with the jet axis, implying that the jet propagates largely out of the disk plane and has limited direct impact on the molecular disk. However, the systematic detection of NE and SW emission across many velocity channels, and across all H$_2$ transitions from low to high excitation, points to the presence of an ongoing shock front roughly perpendicular to the jet axis, likely driven by a bubble inflated by the radio jet. Finally, while outflows are proposed in \citetalias{GarciaBurrillo2009A&A...496...85G} and \citetalias{Ogle2024}, we detect no high-velocity flows ($v>600\,$km\,s$^{-1}$) in either the molecular or ionized gas. Nonetheless, modest outflow velocities are examined in \cref{sec:Kinematics}.

\begin{figure}
    \centering
    \includegraphics[width=0.9\linewidth]{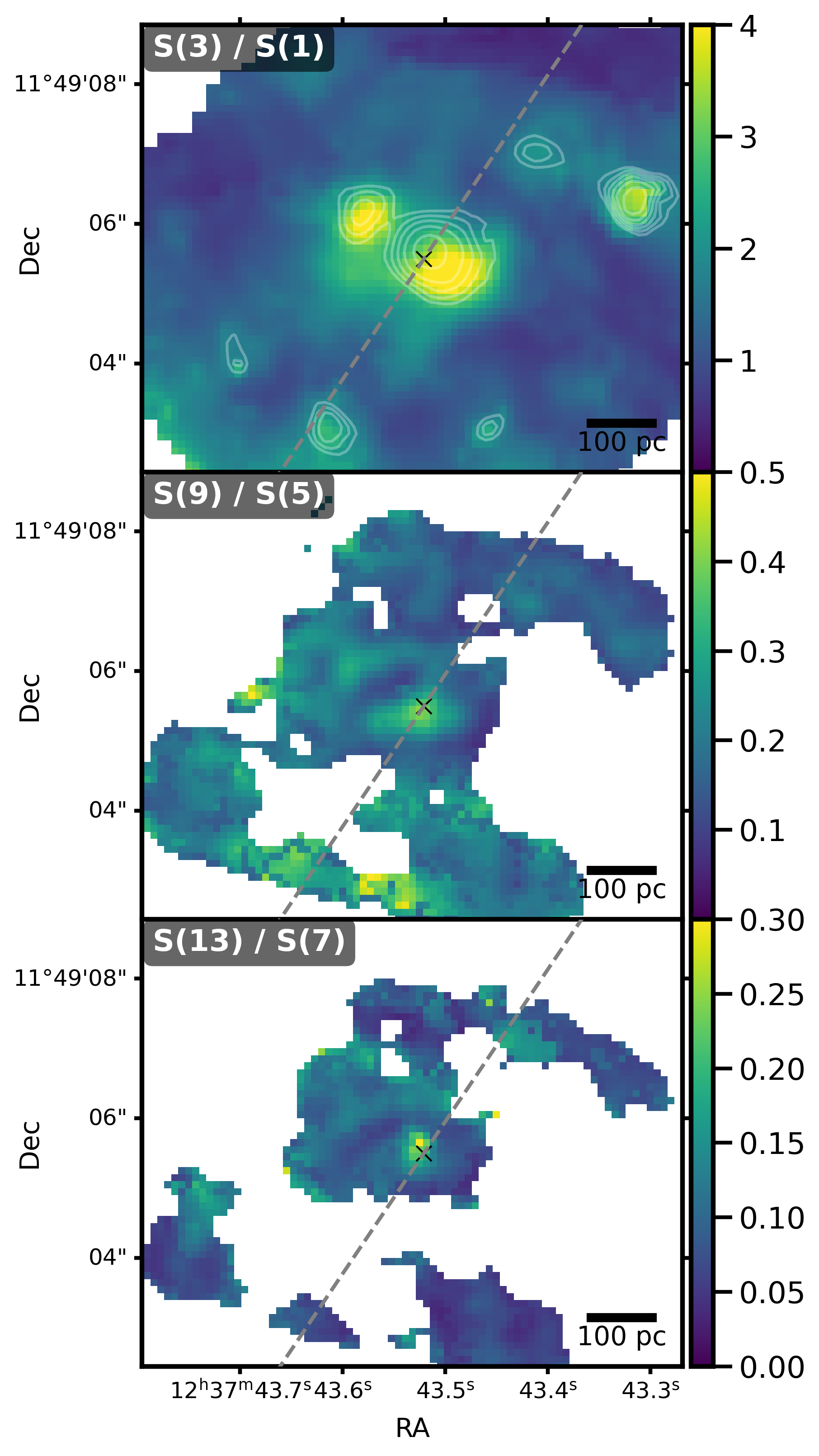}
    \caption{Maps of pure H$_2$ rotational line ratios. \textit{Top:} S(3)/S(1) show elevated values in the dust lanes and in two-lobed structures near the AGN. White contours represent the OPR; the outermost contour corresponds to 3 (expected in LTE at $T$\,$\gtrsim$\,$200$\,K), with inner contours decreasing in steps of 0.25 down to 1.5.
    \textit{Middle:} S(9)/S(5). \textit{Bottom:} S(13)/S(7). These ratios reveal that the higher-$J$ lines trace arc-like structures in both lobes. The dashed gray line indicates the jet axis.}
    \label{fig:s3s1}
\end{figure}

\section{H$_2$ excitation}
\label{sec:Discussion}

Warm H$_2$ in galaxy centers can arise from a variety of excitation mechanisms. In this section, we use spatially resolved diagnostics, excitation modeling, and multiwavelength context to constrain the dominant heating mechanisms and its origin.


\subsection{Pure rotational H$_2$ lines}
\label{subsec:H2_excitation}

Previous work by \citetalias{Ogle2024} showed that the ionized gas in M58’s nucleus is shock-excited, with shock velocities of 170–440 km~s$^{-1}$. Warm molecular gas was inferred to be similarly heated based on an elevated H$_2$/PAH ratio. While such a ratio is useful in integrated analyses, H$_2$ and PAHs often arise from distinct regions. However, spatial variations in the ratio can still reveal where one emission dominates \citep[e.g.,][]{Garcia-Bernete2024A&A...691A.162G}. With JWST’s improved spatial and spectral resolution, we use more physically motivated diagnostics. 

We adopted the S(3)/S(1) ratio as our primary tracer of excitation (\cref{fig:s3s1}), due to its sensitivity to temperature and weak dependence on extinction \citep{Lambrides2019MNRAS.487.1823L}. Since S(3) arises from a higher energy level than S(1), the ratio steeply increases with temperature. In typical SF regions as M58’s spiral arms, S(3)/S(1)\, is approximately \,0.6 \citepalias{Ogle2024}, while in the photodissociation regions (PDRs) of the Orion molecular cloud (OMC), it averages 1.36 \citep{PDR4ALL_2024A&A...687A..86V}. We find S(3)/S(1)\,$>$\,1.5 along the dust lanes and up to $\sim$\,5 near the AGN, forming NE–SW lobes roughly perpendicular to the jet. To probe hotter gas, we used S(9)/S(5) and S(13)/S(7). These ratios range from 0.1–0.3 in the dust lanes and increase to 0.3–0.5 in the lobes. The nucleus indicates that S(13)/S(7)\, is approximately 0.25–0.30, spatially confined within $\sim$20 pc of the AGN, and the NE lobe has lower values. In S(9)/S(5), the NE lobe traces a clear arc of enhanced excitation, marginally seen in S(13)/S(7).

\begin{figure}
    \centering
    \includegraphics[width=0.85\linewidth]{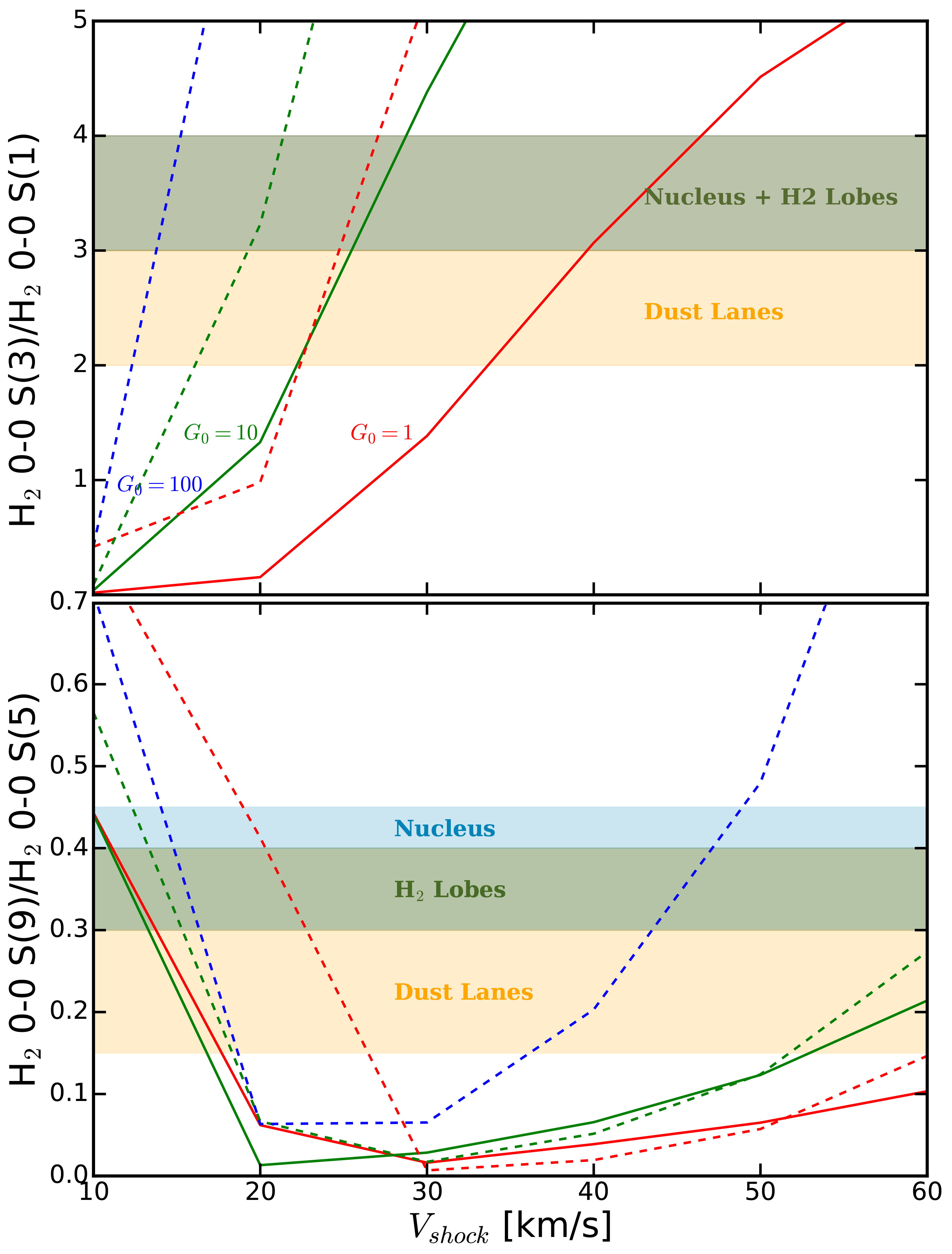}
    \caption{Predicted pure H$_2$ rotational line ratios as a function of shock velocity, based on the C-type shock models of \citet[][]{Kristensen2023A&A...675A..86K}. Two model sets are shown with a magnetic field strength of $B=30\mu$G and pre-shock densities of  $n_H=100$~cm$^{-3}$ (solid lines) and $n_H=1000$~cm$^{-3}$ (dashed lines). Colored curves indicate different strengths of the external UV radiation field $G_0$. Shaded regions mark the range of observed ratios. Constraints on S(13)/S(7) are similar to those for S(9)/S(5).    
    }
    \label{fig:model_rot}
\end{figure}

We compare these results to C‑ and J‑type shock models from \citet{Kristensen2023A&A...675A..86K}, adopting pre-shock densities of $n_{\rm H}$\,$=$\,$10^2$–$10^3$\,cm$^{-3}$ and a magnetic field strength of $B$\,$=$\,$30$\,$\mu$G, consistent with \citetalias{Ogle2024}. 
The external UV radiation field $G_0$ spans a wide range of conditions (1-100), from diffuse clouds to shocked extragalactic environments \citep{Wakelam2017MolAs...9....1W}. Only C-type shocks are able to reproduce the observed H$_2$ line ratios, that primarily depend on shock velocity and density, with only minor sensitivity to external parameters such as $G_0$ or cosmic-ray ionization rates (\cref{fig:model_rot}). The S(3)/S(1) ratio increases steeply with shock velocity across the entire range of parameters. The S(9)/S(5) and S(13)/S(7) ratios match the observed values at low velocities ($\lesssim$10–30\,km\,s$^{-1}$) in all regions and at higher velocities ($\gtrsim$50\,km\,s$^{-1}$) only in the dust lanes. Taken together, these trends consistently point to low-velocity C-type shocks in relatively low-density gas as the dominant H$_2$ excitation mechanism. Alternative mechanisms such as UV pumping preferentially enhance vibrational lines and have minimal impact on low-level $\nu=0$ lines such as S(3). Likewise, variations in the H$_2$ cosmic-ray ionization rate have negligible effect. We further discuss these alternatives for rovibrational excitation in \cref{subsec:rovib}.

\subsection{Ortho-to-para ratio}

\Cref{fig:s3s1} also overlays contours of the H$_2$ ortho-to-para ratio (OPR), defined as the ratio of ortho-H$_2$ (odd rotational levels, $J=1,3,5,\dots$) to para-H$_2$ (even levels, $J=0,2,4,\dots$). We derive the OPR directly from the measured column densities of the ortho transitions (S(1)–S(7)) and the para transitions (S(2)–S(8)) and report it for regions where all eight rotational lines are detected, ensuring a good sampling across a wide temperature range. However, the absence of S(0) in the JWST coverage reduces our sensitivity to any very cold, para-rich reservoir.

In local thermodynamic equilibrium (LTE) at $T$\,$\gtrsim$\,$200$\,K, proton-exchange reactions drive the OPR toward 3 \citep{Reeves1979ZNatA..34..163R}. Across most of the inner kiloparsec, we measure OPR values consistent with this equilibrium, indicating that the H$_2$ has either formed under or relaxed into LTE conditions. However, several compact regions show significantly lower values (OPR\,$\sim$\,$1.5$–2). These include the vicinity of the AGN, encircling the SW lobe and the NE lobe. These regions also exhibit disturbed kinematics (\cref{fig:channels}) and likely trace the aftermath of shock fronts. Additional low-OPR regions are found along the dust lanes. Departures from the equilibrium value can naturally arise when H$_2$ has been dissociated and subsequently reformed on dust grains, yielding a sub-equilibrium OPR. The ratio then evolves gradually toward 3 via proton‐exchange reactions, with characteristic timescales of $\sim10^5$–$10^7\,$yr depending on the gas ionization fraction and density \citep{Wilgenbus2000A&A...356.1010W,Flower2006A&A...449..621F}. The observed sub-equilibrium OPR values therefore likely mark sites of recent H$_2$ reformation or ongoing thermal relaxation following shock passage, consistent also with model predictions from \citet{Kristensen2023A&A...675A..86K}.

\subsection{Modeling excitation diagrams}

To better characterize the warm H$_2$ excitation, we constructed excitation diagrams in each spaxel of the field of view. Plotting the column density per statistical weight, $(N_u/g_u)$, against upper-level energy $E_u$, and assuming optically thin emission and LTE, allowed us to infer the temperature distribution and total column density of the warm H$_2$. While several regions show OPR values below the LTE value, \citet{2010ApJ...724.1193O} showed these deviations have minimal impact on temperatures and masses derived from such diagrams. We used a continuous power-law temperature distribution model introduced by \citet{2016ApJ...830...18T}, which assumes $dN/dT$\,$\propto$\,$T^{-n}$ with $T$\,$\geq$\,$T_l$, where $n$ is the power-law index and $T_l$ is the lower cutoff temperature. Low $T_l$ values (e.g., 50–100~K) are indicative of heating typical of PDRs, while higher values ($\gtrsim$\,200\,K) suggest shock or turbulence-driven heating. For comparison, measurements of M58’s spiral arms from \citetalias{Ogle2024} yield $T_l$\,$\sim$\,$50$~K, consistent with PDRs.

We find that the original implementation from \citet{2016ApJ...830...18T} underpredicts the fluxes of high-excitation lines starting at S(11), as it truncates the temperature distribution at $T$\,$=$\,$2000$\,K. High-J lines are sensitive to hotter gas ($T$\,$>$\,$3000$\,K), which the model excludes. Although two-temperature models can capture this component, they impose an arbitrary separation. To maintain a continuous and physically consistent description, we extended the integration up to the H$_2$ dissociation limit ($T$\,$=$\,$5000$\,K). Because H$_2$ begins to thermally dissociate between $\sim$\,4000–5000\,K \citep{Hollenbach1999RvMP...71..173H}, we introduced an exponential cutoff to suppress unphysical contributions at extreme temperatures. The modified distribution becomes

\begin{equation}
    \label{eq:modtogi}
    \frac{dN}{dT} \propto T^{-n} \, e^{-T / T_{\mathrm{cut}}},
\end{equation}

with $T_{\mathrm{cut}}$\,$=$\,$3500$ K as a soft cutoff and $T_{\mathrm{max}}$\,$=$\,$5000$ K as the upper limit, retaining the flexibility while enforcing a physically motivated suppression near the dissociation threshold.

\begin{figure*}[]
    \centering
    \includegraphics[width=0.95\linewidth]{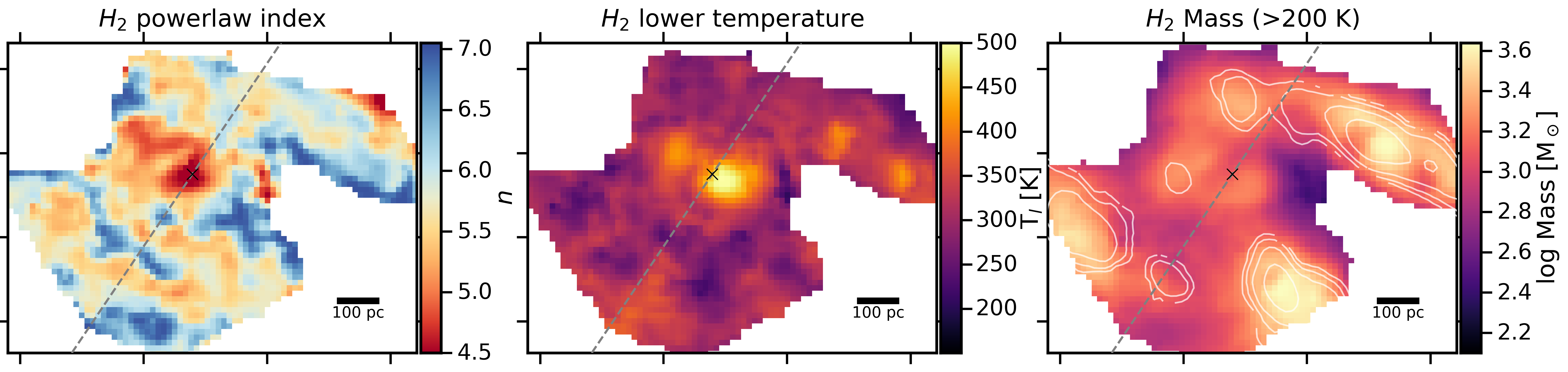}
    \caption{Maps from the modified power-law temperature distribution model from \citet{2016ApJ...830...18T}. Left: Power-law index $n$, with lower values indicating a stronger contribution from hotter gas. Center: Lower temperature $T_l$, contributing to H$_2$ emission. Right: Warm H$_2$ mass per pixel, with each pixel area being $\approx161\,$pc$^2$. White contours trace CO(2–1) emission from cooler molecular gas. The dashed gray line marks the jet axis.}
    \label{fig:physical_ts16}
\end{figure*}

\begin{figure}[]
    \centering
    \includegraphics[width=1\linewidth]{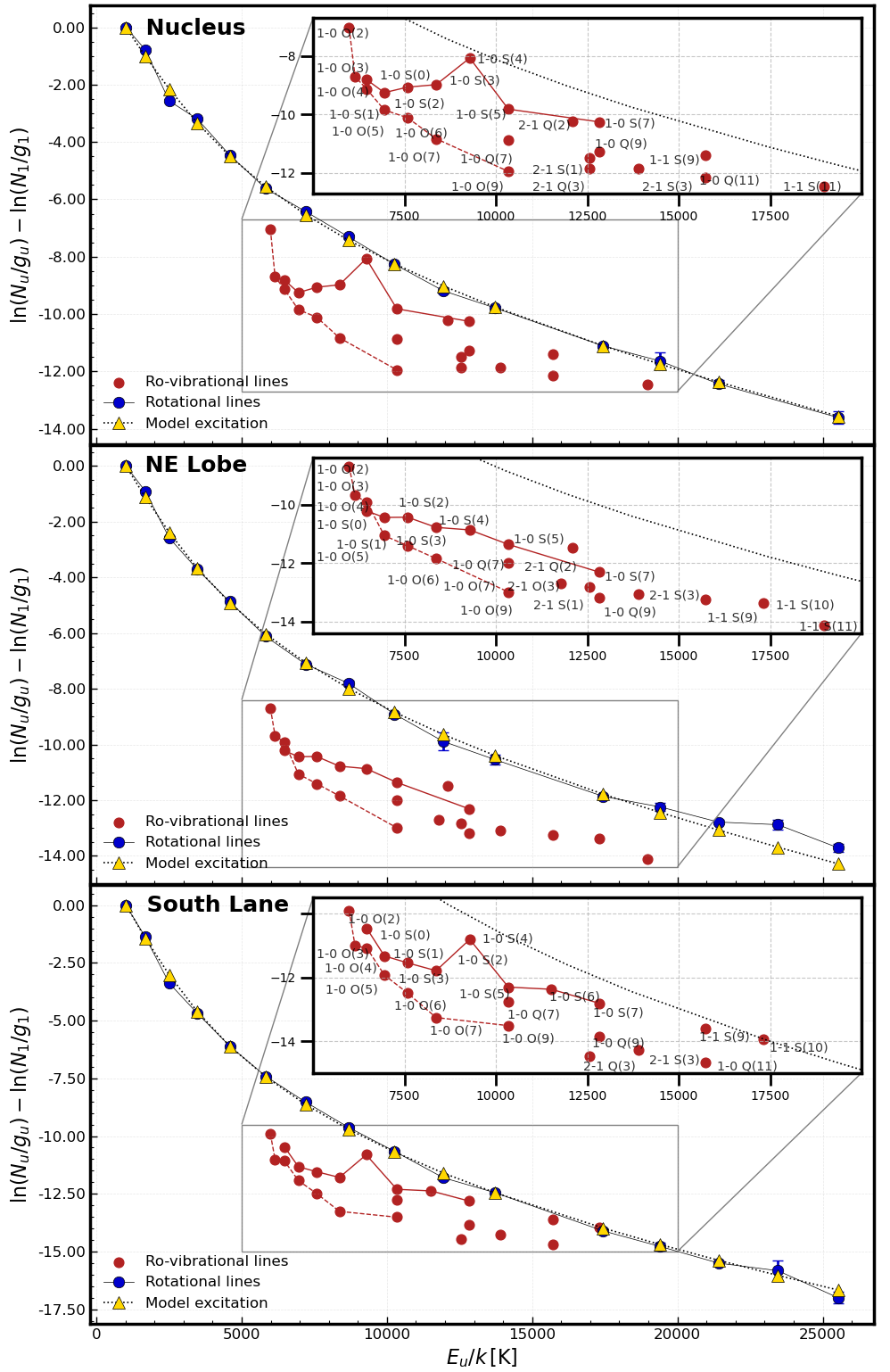}\\
    \caption{H$_2$ excitation ladders for selected regions. Top: Nucleus. Middle: Northeast shock. Bottom: South lane. Pure rotational H$_2$ lines are shown in blue, with best-fit models in yellow triangles. Observed rovibrational lines are plotted in red, with zoomed-in insets displayed in the top corners of each panel. The solid line connects the 1–0 S(1–7) rovibrational ladder, while the dashed line connects the 1–0 O(2–9) ladder.}
    \label{fig:togi}
\end{figure}

\Cref{fig:physical_ts16} shows the fitted parameters across the field, and example excitation diagrams for selected regions appear in \cref{fig:togi}. The power-law index $n$ indicates the steepness of the temperature distribution: high $n$ implies that cooler gas dominates; low $n$ implies that hot gas dominates. Along the dust lanes, we find high $n$ values (up to $\sim$6.5), steep temperature distributions, and $T_l$\,$\sim$\,300–350~K. In contrast, regions between the nucleus and lanes show flatter distributions ($n$\,$\sim$\,4–5) and lower $T_l$\,$\sim$\,$200$\,K, with the nucleus having the flattest slope ($n$\,$\approx$\,$3.9$) and highest cutoff temperature, signaling a dominant hot component likely heated by AGN-driven processes. The NE–SW lobes also reach $T$\,$\sim$\,$500$\,K, indicating a large warm gas contribution. Note that the $T_l$ parameter in the model from \cref{fig:togi} represents a lower bound on the gas temperature.

We converted the fitted temperature distributions into warm H$_2$ masses ($T$\,$>$\,$200$\,K), following \citet{2016ApJ...830...18T}, using the spaxel area of MRS CH1 (0.0168 arcsec$^2$ or $\sim$161 pc$^2$). Most warm gas lies in the dust lanes, and a clear disruption in the southern lane is traced by H$_2$, as also seen in CO \citepalias{GarciaBurrillo2009A&A...496...85G} and reflected in kinematics (\cref{fig:mom_miri}). Despite their compact size (each lobe spanning roughly 100\,pc $\times$ 50pc), the two molecular lobes contain a substantial amount of warm H$_2$, each one with approximately $5 \times 10^4$~M$\odot$ of warm H$_2$.

\subsection{Rovibrational H$_2$ lines}
\label{subsec:rovib}

The excitation diagrams in \Cref{fig:togi} show clear decoupling between the pure rotational and rovibrational H$_2$ ladders across regions. Were the gas in LTE fit by a single temperature, both sets of transitions would fall along the same curve in a Boltzmann plot. Instead, the $\nu=1$–0 lines (both S- and O-branches) lie systematically below the rotational trend. High-energy rovibrational lines (e.g., 1–0 S(7), $E_u/k\gtrsim10^4\,$K) align more closely with the rotational ladder, whereas low-energy lines (e.g., 1–0 S(1)) are offset by up to two dex. This curvature, in which the excitation temperature increases with $E_u$, is characteristic of shock-dominated environments where higher-$J$ and rovibrational transitions preferentially trace hotter gas \citep{Kristensen2023A&A...675A..86K}.

This misalignment has been observed in a variety of galaxies: for instance, NGC 1266 requires mixed PDR and AGN contributions to reproduce its H$_2$ ladder \citep{Otter2024ApJ...975..142O}, and NGC 3256 combines slow shocks and X-ray heating to reach rovibrational temperatures near 2000 K \citep{Emonts2014A&A...572A..40E}. Similarly, protostellar outflows show LTE-like pure rotational lines alongside sub-thermal rovibrational levels in nonequilibrium OPR regions \citep{Vleugels2025A&A...695A.145V}. In M58, the divergence between S and O branches persists even where OPR$\approx3$, confirming that a simple OPR effect cannot account for the offset.

\begin{figure*}[t]
    \centering
    \includegraphics[width=0.75\linewidth]{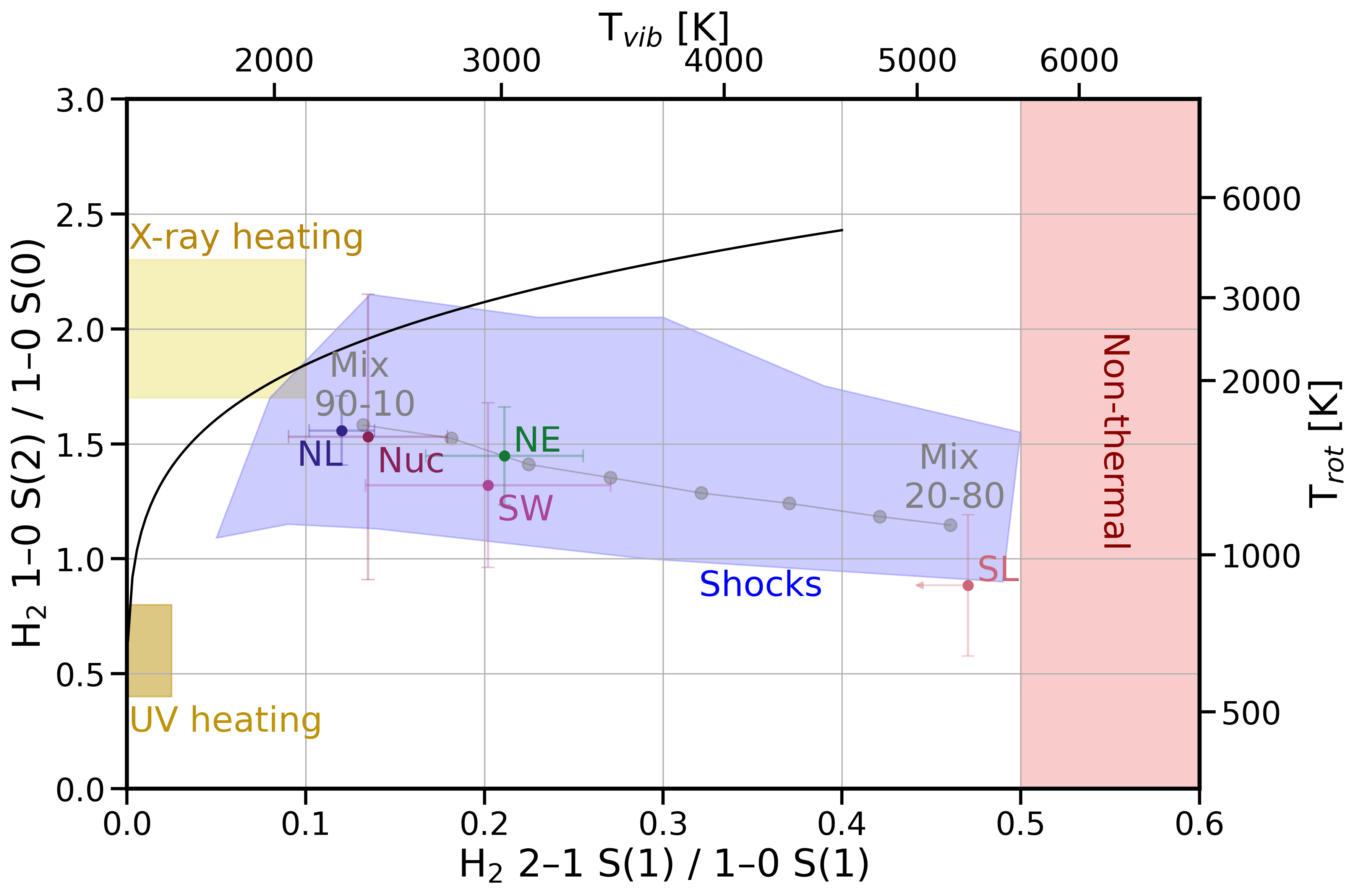}
    \caption{H$_2$ 2–1 S(1)/1–0 S(1) vs. 1–0 S(2)/1–0 S(0) emission-line ratios for selected regions of M58. Colored markers indicate the nucleus (Nuc), north and south lane (NL-SL), and the northeast and southwest H$_2$ lobes (NE-SW). The solid black curve marks the predicted trend for pure thermal excitation. Shaded regions represent the domains of alternative excitation mechanisms: UV thermal excitation \citep{Sternberg1989ApJ...347..863S}, X-ray heating \citep{Draine1990ApJ...363..464D}, shocks \citep{Kristensen2023A&A...675A..86K}, and nonthermal UV fluorescence \citep{Black1987ApJ...322..412B}. The gray curve shows mixed thermal and fluorescent excitation models, decreasing from 90\% thermal (left) in 10\% steps. The top and right axes indicate the vibrational and rotational temperatures (T$_{\rm vib}$, T$_{\rm rot}$) respectively.}
    \label{fig:grid}
\end{figure*}

The systematic underpopulation of the $\nu\ge1$ transitions relative to the pure rotational ladder demonstrates that a single-temperature model cannot account for both rotational and vibrational lines. Although a temperature distribution of $T_u=150$–500 K reproduces the pure rotational data, it overestimates the $\nu=1$–0 rovibrational line intensities. This discrepancy arises because rovibrational levels have critical densities on the order of $\sim10^6\,$cm$^{-3}$ (e.g., $v=1$, $J=3$ at $T\gtrsim2000\,$K), whereas low-$J$ rotational lines require densities of only $\lesssim10^3\,$cm$^{-3}$ (e.g., S(1)). In regions where $n_{\rm H}<n_{\rm crit}$, collisions cannot sustain thermal level populations, leading to sub-thermal excitation and correspondingly weaker rovibrational emission. This sub-thermal excitation naturally produces the flattening of the 1–0 S(1-7) ladder, reflecting variations in gas density and excitation depth across the field. However, additional excitation mechanisms may also contribute. The rovibrational lines may arise from gas subject to different shock velocities or irradiated by energetic photons. In particular, photon-driven processes such as UV or X-ray pumping can preferentially populate high vibrational states, modifying the ladder independently of the thermal background.

Another useful diagnostic compares pure rotational with rovibrational lines, such as S(9)/1–0 S(5) and S(13)/1–0 O(7), which can be readily derived from the fluxes listed in \cref{table:h2}. These ratios probe different excitation regimes, revealing variations in shock strength and post-shock cooling. The two lanes and the NE region exhibit S(9) dominance ($\sim$1.2). Only the nucleus exhibits an S(9)/1–0 S(5) of $\sim$0.85, implying hotter conditions with more efficient rovibrational excitation. The S(13)/1–0 O(7) lines are at similar wavelengths, minimizing extinction effects, and trace density in partially thermalized gas. The critical density of S(13) ($n_\mathrm{crit}\sim10^6\,$cm$^{-3}$) is much higher than 1–0 O(7). In the OMC, the ratio is uniformly $\sim$1.05, indicating full thermalization at $n_\mathrm{crit}$, while in extreme shocked regions it can rise to $\sim$2.3 due to overpopulated high-$J$ levels under sub-thermal conditions \citep{Brand1989MNRAS.236..929B}. In M58, the nucleus reaches peak values of $\sim$1.95, consistent with high temperatures and sub-thermal excitation, while values $\lesssim$1.3 in the dust lanes indicate denser, near-LTE conditions. Together, these trends support a picture of cooler dust lanes and hotter, less dense lobes, with local density variations consistent with turbulent post-shock structures.


\Cref{fig:grid} presents the 2–1 S(1)/1–0 S(1) versus 1–0 S(2)/1–0 S(0) diagram \citep[e.g.,][]{Mouri1994ApJ...427..777M, Rodriguez2004A&A...425..457R, Mazzalay2013MNRAS.428.2389M}, which jointly constrains the vibrational ($T_{\rm vib}$) and rotational ($T_{\rm rot}$) excitation temperatures of the gas.\footnote{See equations 1 and 2 in \citet{Mazzalay2013MNRAS.428.2389M}.} In LTE and under purely thermal excitation, the line ratios fall along a curve where $T_{\rm vib}$\,$\approx$\,$T_{\rm rot}$. Deviations can be explained by model predictions for UV and/or X-ray heating, shocks, and when $T_{\rm vib}$\,$>$\,$T_{\rm rot}$, indicating the presence of nonthermal excitation. Although the 2–1 S(1)/1–0 S(1) ratio is often used to identify UV pumped gas, it is sensitive to $T_{\rm vib}$ and may reflect vibrational thermalization under shock conditions. High $T_{\rm vib}$ values have been observed in dense shocked clouds such as the OMC \citep{Habart2005SSRv..119...71H}, but in M58 the conditions appear more consistent with a lower-density regime, where collisions are less effective at thermalizing vibrational levels. Under these circumstances, the 2–1/1–0 ratio remains a valid diagnostic of the relative contribution of the thermal versus nonthermal contribution. With all regions showing values below 0.5, we can confidently rule out pure nonthermal fluorescence in M58. Instead, all regions are broadly consistent with collisional excitation driven by shocks. The nucleus and north lane lie closest to the pure thermal curve, consistent with $\sim$90\% thermal and $\sim$10\% nonthermal excitation. The NE and SW lobes shift toward larger nonthermal fractions ($\sim$20-30\%). The south lane has weak 2–1 S(1) emission and cannot be fully constrained. These mixed-excitation fractions align with surrounding physical conditions: shocks dominate in the central regions and along the north lane, while the lobes, located between the AGN and the UV ring, experience enhanced radiative excitation. This behavior echoes findings in other LINER and Seyfert 2 galaxies, which typically cluster near the thermal curve with modest nonthermal offsets. This is in contrast to Seyfert 1 and starburst systems, which show stronger fluorescent signatures \citep{Riffel2013MNRAS.430.2002R,Otter2024ApJ...975..142O}. 

\subsection{Origin of H$_2$ excitation}
\label{subsec:H2_origin}

We have shown that the pure rotational lines are nearly fully thermal, while the rovibrational lines contain only a 10–30\% nonthermal component. We now assess the physical drivers responsible for these excitation processes.

\subsubsection{Thermal dominance}

\begin{figure*}
    \centering
    \includegraphics[width=0.75\linewidth]{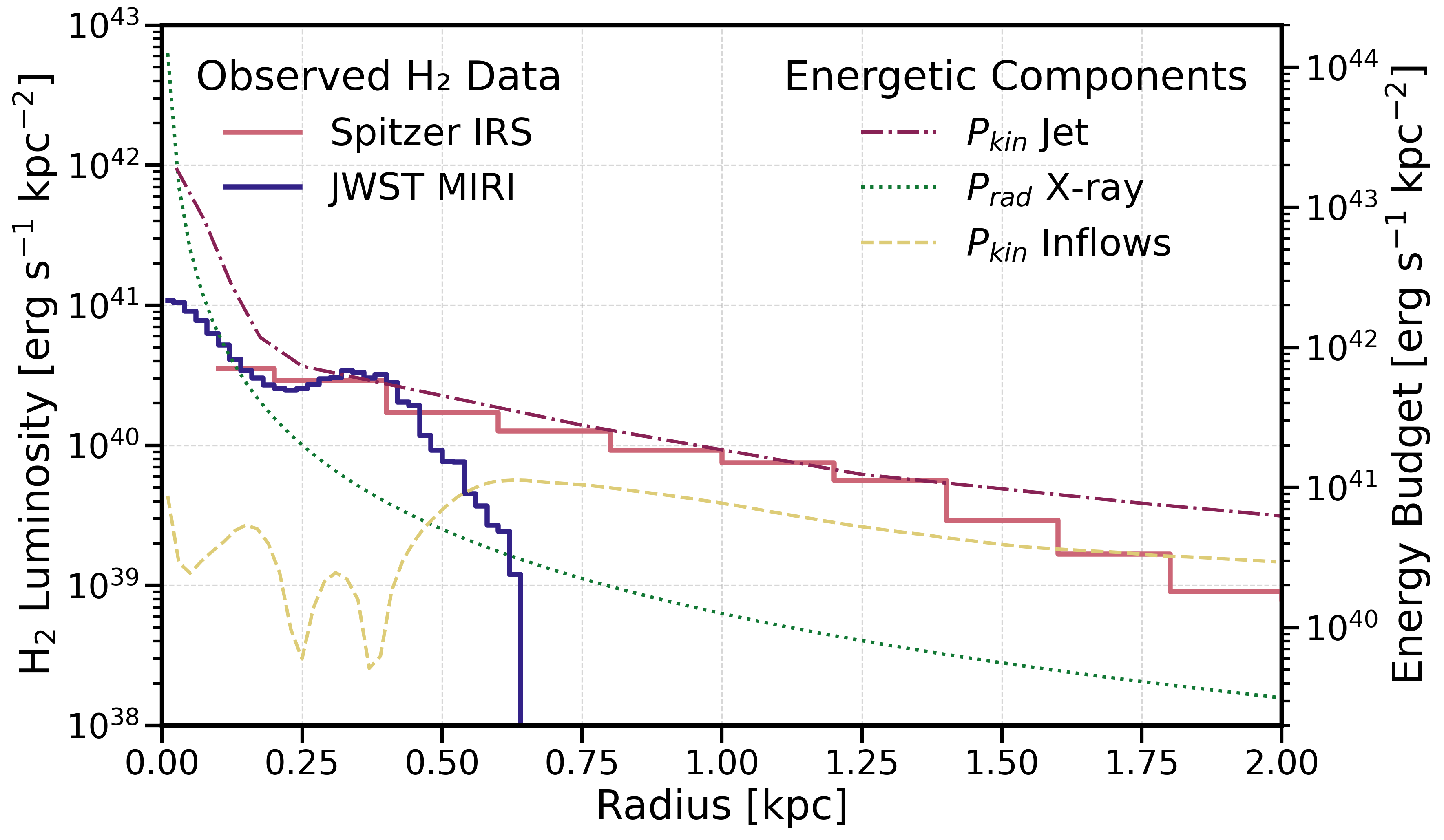}
    \caption{Radial profile of H$_2$ luminosity surface density compared to various energy input mechanisms. Spitzer IRS and JWST MIRI H$_2$ S(3) measurements are scaled to match the total S(1)–S(5) luminosity. The JWST-derived surface density drops beyond 0.5 kpc due to the limited field of view. Energy surface densities, assuming an overall coupling efficiency of $\eta = 5\%$, are shown for X-ray radiation pressure and kinetic power from cloud–cloud collisions and from the radio jet.}
    \label{fig:energy budget}
\end{figure*}

Since low-lying pure-rotational lines dominate thermal H$_2$ cooling, \cref{fig:energy budget} compares the radial profile of the rotational luminosity, scaled from S(3) to the total S(1)–S(5) emission\footnote{S(3) contributes $\sim$35\% of the summed S(1)–S(5) luminosity measured by Spitzer over the central 2 kpc.} to the energy available from three candidate mechanisms, each assuming a conservative 5\% coupling efficiency. We estimated the radio jet’s mechanical power, using VLA 5 GHz and LOFAR 144 MHz maps and the scaling relations of \citet{2010ApJ...720.1066C}\footnote{These were calibrated at 1.4\,GHz and 323\,MHz; we extrapolated assuming a typical synchrotron spectral index of $\alpha$\,$=$\,$-0.8$}. X-ray heating of the AGN was evaluated using the observed luminosity accounting for $r^{-2}$ geometric dilution. The cloud–cloud collision power was calculated via $P_{\rm kin}=\tfrac12\,\dot M\,v^2$, where $\dot M$ represents the inflow and outflow rates from \citetalias{GarciaBurrillo2009A&A...496...85G}, assuming $v$\,$=$\,$150$\,km\,s$^{-1}$ for cloud encounters. 

Even at 5\% efficiency, the jet’s kinetic power exceeds the H$_2$ surface brightness across the full 2 kpc. By contrast, X-ray heating is limited to the inner $\sim$100\,pc and cloud–cloud collisions can sustain the emission only beyond $\sim$1.5 kpc. Even combined, these two processes cannot account for the prominent H$_2$ bump at 0.3–0.4 kpc, coincident with the onset of the dust lanes. Despite the simplified assumption, the radio jet alone provides sufficient energy to power the thermal H$_2$ emission at all scales. Other processes, even if locally relevant, would require implausibly high coupling efficiencies: XDR heating typically operates at $\lesssim$1\% \citep{2010ApJ...724.1193O}, and while cloud–cloud collisions can be efficient in extreme interacting systems \citep[e.g., Stephan’s Quintet;][]{Guillard2009A&A...502..515G}, there is no evidence for recent mergers or interactions in M58. We did not explore other origins of the shocks, such as supernova-driven shocks, given the lack of evidence for a past intense starburst. Likewise, the morphology of the shocked region does not support external drivers such as cluster ram pressure \citep[e.g., NGC 4569;][]{Boselli2016A&A...587A..68B}.

\subsubsection{Nonthermal contribution}
\label{subsec:H2_origin_rob}

In the nucleus, evidence suggests a nonthermal excitation of $\sim$10\%. Although our excitation grids model nonthermal excitation solely via UV fluorescence, other processes (e.g., cosmic rays and XDRs) could similarly populate the $v=2$ level.

The observed nuclear 2–1 S(1) luminosity is $L_{\rm 2-1 S(1)} \approx 3.25 \times 10^{37}\,\mathrm{erg\,s^{-1}}$. This line arises from the level ($v=2$, $J=3$), typically populated via radiative cascades following UV pumping through electronic transitions (e.g., B$^1\Sigma_u^+$ $\rightarrow$ X$^1\Sigma_g^+$). The probability that an absorbed UV photon ultimately leads to population in $v=2$ is set by the Franck–Condon factors, which for $v=2$ are $\sim$0.15 \citep{Spindler1969JQSRT...9.1041S}. Adopting a conservative 10\% cascade efficiency \citep{Sternberg2014ApJ...790...10S} obtain an overall UV-to-2–1 S(1) conversion efficiency of $\eta\approx0.015$. From LLAGN models, the intrinsic far-UV luminosity of the AGN is $L_{110\mathrm{nm}}$\,$\approx$\,$6.68$\,$\times$\,$10^{33}\,\mathrm{erg\,s^{-1}\,nm^{-1}}$ \citep{Lopez2024A&A...692A.209L}, corresponding to a photon rate of $\dot{N}_{\rm UV}$\,$\sim$\,$1.38$\,$\times$\,$10^{42}$~s$^{-1}$. Thus, the predicted 2-1 S(1) luminosity is $L_{\rm pred}$\,$=$\,$\dot{N}_{\rm UV} \times \eta \times h\nu$\,$\approx$\,$1.83 \times 10^{28}~\text{erg~s}^{-1}$, nine orders of magnitude below the observed value. Using HST fluxes that include UV emission from young stars yields similarly low results: extrapolating from 250\,nm \citep{Maoz2005ApJ...625..699M}, we estimate $L_{110\mathrm{nm}}$\,$\sim$\,$1.7\times10^{39}\,\mathrm{erg\,s^{-1}\,nm^{-1}}$, implying $\dot{N}_{\rm UV}$\,$\sim$\,$3.5\times10^{47}\,\mathrm{s^{-1}}$ and $L_{\rm pred}$\,$\sim$\,$4.7\times10^{33}\,\mathrm{erg\,s^{-1}}$. Even assuming perfect trapping, UV fluorescence falls orders of magnitude short. We therefore conclude that UV pumping is negligible for the observed high-$v$ H$_2$ emission in the nucleus.

X-rays penetrate deeper into molecular clouds than UV photons, generating fast photoelectrons that collisionally excite H$_2$ to high vibrational levels. Given the intrinsic AGN X-ray luminosity of $L_X$\,$\sim$\,$1.6\times10^{41}\,$erg s$^{-1}$ and a clumpy medium with $n$\,$\sim$\,$10^3$\,cm$^{-3}$, soft X-rays can be absorbed within a few parsecs, and XDR models predict $T$\,$\sim$\,$10^3$\,K gas and efficient $v=2$ excitation \citep{Draine1996ApJ...468..269D}. Accounting for the approximately 10\% nonthermal 2–1 S(1) emission requires $\sim$\,$760\,M\odot$ of warm H$_2$, only approximately 3\% of the total $T>200$\,K gas we infer in the nucleus. Similar warm gas fractions and X-ray-driven H$_2$ rovibrational excitation have been observed in other AGN \citep[e.g.,][]{Riffel2013MNRAS.430.2002R,Storchi2009MNRAS.394.1148S}, making X-ray pumping a viable explanation for the nuclear nonthermal component.

Cosmic rays can excite the H$_2$, primarily populating the $v=1$ vibrational level \citep{Padovani2022A&A...658A.189P}; however, producing the observed emission would require extremely high ionization rates. Adopting a nonthermal 1–0 S(1) luminosity of $2.41\times10^{37}\,$erg s$^{-1}$ (10\% of the total) and a warm H$_2$ reservoir of $\sim2.5\times10^4\,M_\odot$ at $T>200\,$K, the required ionization rate is $\sim$\,$10^{-10}$–$10^{-9}$\,s$^{-1}$, depending on the assumed photon yield per ionization ($10^{-3}$–$10^{-2}$). These values are orders of magnitude higher than typical Galactic molecular cloud rates ($10^{-16}$–$10^{-15}$\,s$^{-1}$ \citealt{Indriolo2012ApJ...745...91I,Neufeld2017ApJ...845..163N}) and even exceed the highest rates seen in the Galactic Center or strong shocks ($\sim$\,$10^{-14}$–$10^{-13}$\,s$^{-1}$; \citealt{Padovani2018A&A...614A.111P,Indriolo2015ApJ...800...40I}). In AGN environments, molecular ion studies typically infer rates no higher than $\sim$\,$10^{-12}$–$10^{-13}$\,s$^{-1}$ \citep[e.g.,][]{GonzalesAlfonso2018ApJ...857...66G,Holdship2022ApJ...931...89H}. Cosmic rays may contribute in localized, turbulent subregions but cannot account for the bulk of the nonthermal H$_2$ luminosity.

\section{Dynamical imprint of LLAGN feedback}
\label{sec:Kinematics}

\begin{figure*}
    \centering
    \includegraphics[width=0.95\linewidth]{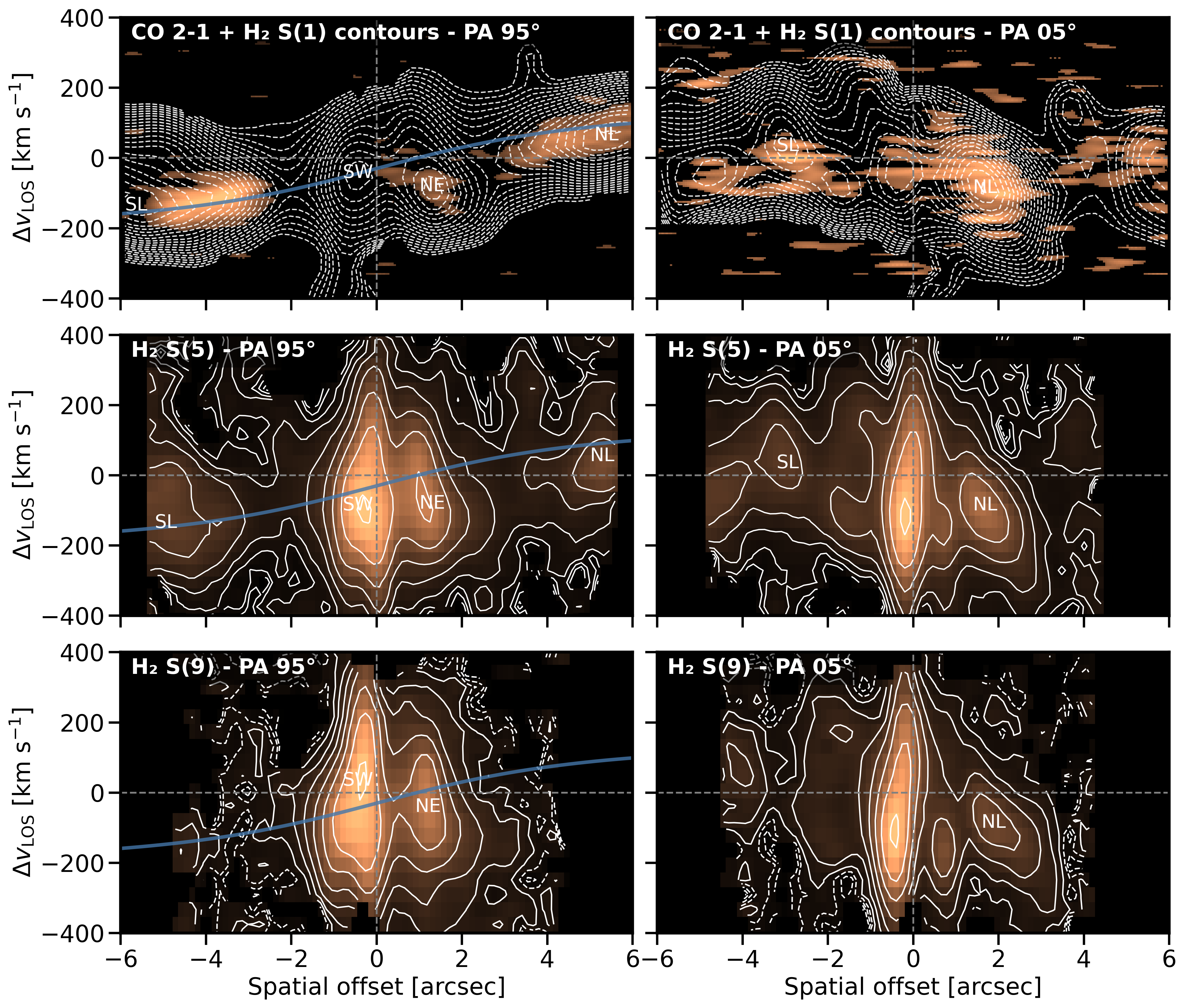}
    \caption{PV diagrams of molecular gas emission along the major (PA = 95°) and minor (PA = 5°) axes of the galaxy. Top row: CO(2–1) emission from \citetalias{GarciaBurrillo2009A&A...496...85G} as a color map, with white contours of the H$_2$ S(1) line overlaid. A velocity shift of 30 km\,s$^{-1}$ was applied to the CO data to match our systemic velocity. Second and third rows: H$_2$ S(5) and S(9) emission lines, respectively. In all major-axis panels, the blue line marks the best-fit CO rotation curve. Contours begin at 10$\sigma$.}
    \label{fig:pv}
\end{figure*}

\begin{figure*}
    \centering
    \includegraphics[width=1.0\linewidth]{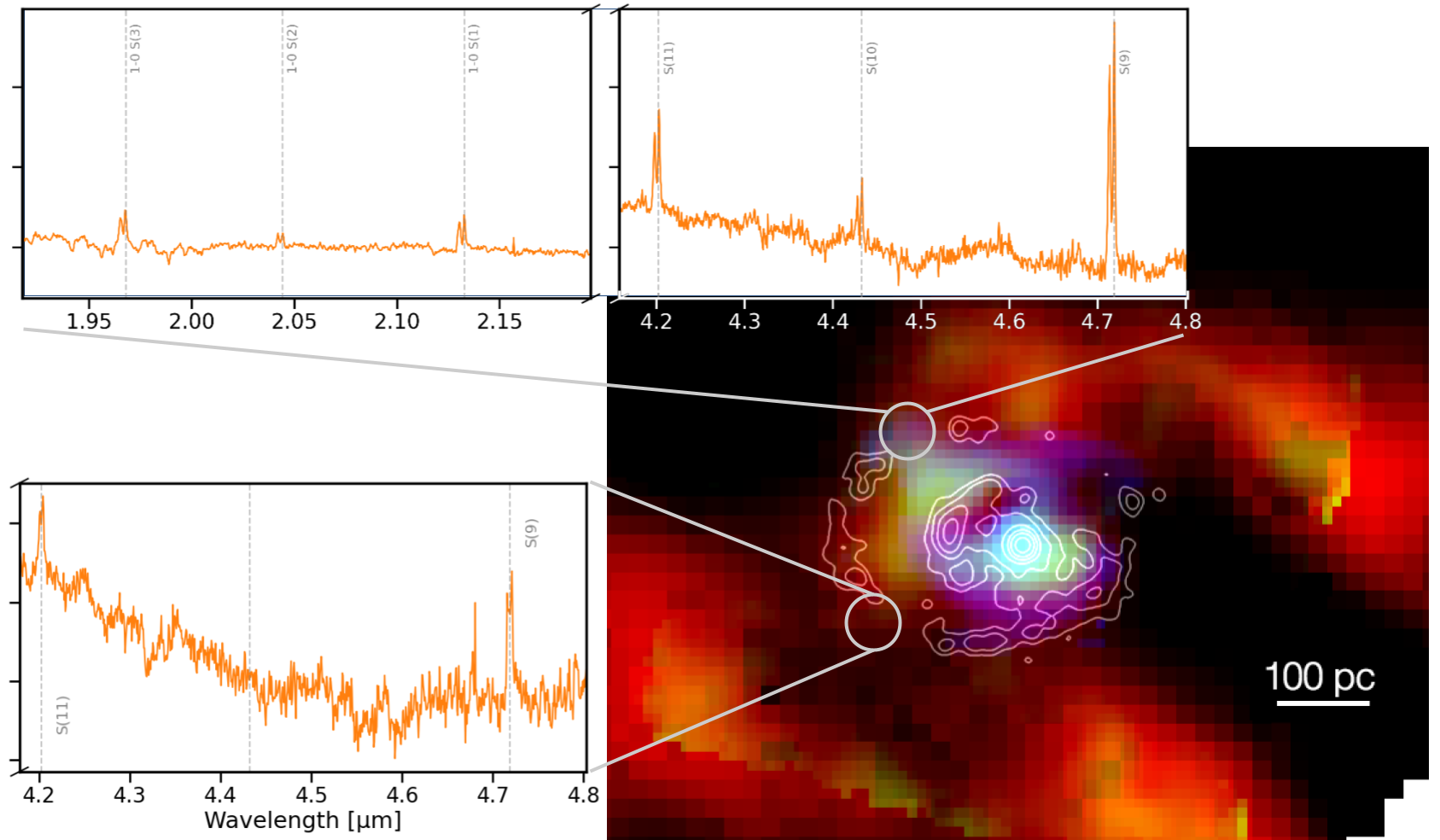}
    \caption{Color composite: H$_2$ S(1) (red), H$_2$ 1–0 S(1) (green), and Pa$\alpha$ (blue) with UV contours. The spectra show the double-horned profiles of the high excitation H$_2$ lines.}
    \label{fig:vortex}
\end{figure*}

As shown in \cref{sec:3}, the velocity fields reveal a large-scale pattern consistent with rotation across the central kiloparsec, with projected velocities ranging from –200 to +200\,km\,s$^{-1}$. This disk-like structure, which agrees with the CO kinematics reported in \citetalias{GarciaBurrillo2009A&A...496...85G}, indicates that AGN feedback in M58 does not strongly disrupt the global velocity field, unlike the high-velocity molecular outflows seen in more luminous AGN \citep[e.g.,][]{Rupke2013ApJ...775L..15R,Ramos2019MNRAS.487L..18R}. Instead, the feedback more subtly imprints on the molecular phase, primarily through thermal excitation, as seen in other jetted systems such as the Teacup galaxy \citep{Zanchettin2025A&A...695A.185Z}. Nonetheless, localized deviations from rotation are evident in \cref{fig:channels}. \citetalias{GarciaBurrillo2009A&A...496...85G} confirmed the presence of noncircular motions and streaming features in the inner disk of M58, complicating the interpretation of disturbed gas as purely outflow-driven.

In the central $\sim$50\,pc, the velocity dispersion increases reaching several hundred kilometers per second. This broadening is most pronounced in the ionized gas, where [Ar\,\textsc{ii}] reaches $\sim$400\,km\,s$^{-1}$ and rovibrational H$_2$ ($\nu > 1$) lines approach 200\,km\,s$^{-1}$. In contrast, pure rotational H$_2$ lines exhibit narrower widths of 100–150\,km\,s$^{-1}$, suggesting that the most energetic components trace turbulent substructures that are locally decoupled from the galactic disk.

Position–velocity (PV) diagrams (\cref{fig:pv}) along the galaxy’s major and minor axes, with position angles (PA) of 95$^\circ$ and 5$^\circ$, respectively, trace warm H$_2$ emission associated with the CO-defined dust lanes (SL and NL). The emission extends to $\sim$400–600,pc along the major axis and $\sim$200–300,pc along the minor axis. The major-axis lanes follow the expected rotation curve, but the minor-axis PV diagram shows systematic velocity offsets, consistent with radial flows or in-plane outflows. The so-called outer arc, associated with a bar-driven resonance \citepalias{GarciaBurrillo2009A&A...496...85G}, lies just beyond the JWST field of view. Notably, the northern lane lies closer to the nucleus and exhibits a significantly higher velocity dispersion than the southern one. This asymmetry may reflect uneven inflow rates, interactions with the AGN jet, or the presence of a dynamically heated gas bridge connecting the nucleus to the northern lane. The H$_2$ S(1) line reveals fainter, more extended emission in regions where CO is weak or absent, especially in shocked zones. The NE complex departs from the expected rotation and is mirrored by a southwestern SW structure lacking a CO counterpart, together suggesting a bipolar morphology. \citetalias{GarciaBurrillo2009A&A...496...85G} placed the NE structure in a kinematically forbidden region for pure-rotational motion, supporting its interpretation as an outflow. This component appears in H$_2$ S(1) and becomes more prominent in S(5) and S(9), where both NE and SW features extend further in velocity but the NE approaches the rotational curve. In S(5) and S(9), the NE and SW regions extend to higher velocity dispersions, although the NE structure aligns more closely with the underlying rotational curve.

Along the minor axis, the H$_2$ S(5) line displays a broad component centered at -100\,km\,s$^{-1}$, extending spatially from -200 to +200\,pc. At higher excitation (S(9)), the southern lane fades significantly, whereas the northern complex remains bright, suggesting asymmetric energy injection or excitation. While jet signatures are subtle, extended northern emission seen in VLA maps may trace gas entrained in a weak jet-driven outflow. These features could correspond to out-of-plane motions, with de-projected velocities of 120–170\,km\,s$^{-1}$. The NE complex shows weak associated radio emission but otherwise resembles an in-plane shock, mirrored by the SW structure. Together, they delineate a C-type shock front tracing the boundary of an expanding bubble driven by jet feedback. The morphology seen in the channel maps, combined with the estimated molecular gas masses and the alignment of higher-excitation lines with the disk rotation, all support the interpretation that these structures lie within the galactic disk rather than above or below it. This indicates that the bulk of the shocked gas remains within the disk, even if the jet escapes the galaxy plane. De-projected velocities of the H$_2$ S(1) line in these regions typically reach $\sim$170\,km\,s$^{-1}$, while higher-excitation lines such as S(9) extend up to $\sim$340\,km\,s$^{-1}$.

Outside the jet axis, double-horned line profiles in the warm H$_2$ emission highlight the disrupted nature of the shock’s leading edge, as shown in \cref{fig:vortex}. They appear in regions that spatially align with interruptions in the SF ring, where high-excitation H$_2$ transitions coincide with non-UV emission. Spectra in two of these regions within a $\sim$30 pc aperture reveal two symmetric, narrow velocity components with a separation of $\sim$100 km\,s$^{-1}$, and both showing similarly high excitation levels. Further evidence of stratified feedback emerges from the spatial decoupling between warm H$_2$, Pa$\alpha$, and UV. The UV emission tracing the SF ring lies just outside the warm H$_2$ in the outflowing NE structure, with a similar configuration observed in the SW, where the ring borders the outer edge of the SW shock. This spatial sequence is consistent with a scenario in which mechanical energy injected by the expanding bubble driven by the jet is propelling H$_2$ outflows and displacing large gas masses that compress the ISM, enabling SF to proceed relatively undisturbed in a surrounding ring. Rayleigh–Taylor instabilities may also arise where a low-density shock collides with denser material from the SF ring, likely producing the observed double-horned H$_2$ profiles.

In the southern part of the SF ring, a partial decoupling of Pa$\alpha$ and UV emission suggests that the most massive stars ($\gtrsim$20\,M$\odot$) have already exploded as supernovae and dispersed their surrounding H\textsc{ii} regions. The remaining UV could originate from intermediate-mass stars (3–10\,M$\odot$), which emit nonionizing UV radiation for up to $\sim$30\,Myr, implying that the last SF burst ended 8–30\,Myr ago. Sub-equilibrium OPR values point to recent shock activity with estimated ages of 0.1–1\,Myr \citep{Neufeld2006ApJ...649..816N}. Taken together, the OPR, UV and Pa$\alpha$ distributions may trace distinct but complementary timescales. Assuming similar shock velocities, the travel time to 200\,pc would be $\sim$1\,Myr, while reaching kiloparsec scales would require $\sim$10\,Myr. This suggests that the kpc-scale cocoon observed by \citetalias{Ogle2024} is consistent in age with the last SF burst. In contrast, the warm H$_2$ excitation seen in the NE and SW traces a more recent ($\sim$1\,Myr) shock event, likely resulting from ongoing interaction between the jet and the ISM. A full analysis of the ionized gas and the PAH emission will be presented in a forthcoming work to characterize the multiphase interplay and impact on SF.

Overall, LLAGN feedback in M58 does not drive fast winds or expel large gas reservoirs. Instead, it acts subtly, disturbing the inner 200\,pc through C-type shocks and turbulence, with warm H$_2$ outflows reaching $\sim$170\,km\,s$^{-1}$, while the large-scale rotating disk is just thermally heated. Other observational studies support this scenario. An excess of warm H$_2$ is commonly seen near AGN \citep{Lambrides2019MNRAS.487.1823L}. In some cases, such as NGC 3100, the jet mildly perturbs the gas kinematics \citep{Ruffa2022MNRAS.510.4485R}. Broader samples show that, beyond a certain radius, jets typically have minimal dynamical impact on the ISM \citep{Ayubinia2023ApJ...954...27A,Kukreti2025A&A...698A..99K}. In low-Eddington phases, molecular gas builds up around the nucleus, and only jets aligned with the disk can remove it efficiently \citep{Garcia2024A&A...689A.347G}.

Ultimately, M58 illustrates the long-term impact of low-power AGN feedback, primarily driven by an off-plane radio jet and, in second order, by X-ray photons from the ADAF. Rather than clearing the ISM, the AGN regulates it by heating the inner disk via C-type shocks and quenching SF up to kiloparsec scales with non-dynamical coupling. In regions where shocks impact dense gas directly, we observe mildly disturbed kinematics and excitation conditions that resemble the effects of more powerful AGN, albeit on smaller scales. These findings highlight the cumulative impact of sustained, low-level LLAGN feedback in shaping both the ISM structure and the SF potential of galaxies.


\section{Summary and conclusions}
\label{sec:5}

JWST's unprecedented sensitivity opens a new window into the subtle but significant impact of LLAGN feedback. In this study, we presented the most detailed near and mid-IR view to date of the inner kiloparsec of M58, revealing the physical conditions, excitation mechanisms, and kinematics of warm H$_2$ in the vicinity of a weak AGN with a low-power radio jet. Our main findings can be summarized as follows:

\begin{enumerate}
    \item We detect pure bright H$_2$ rotational lines (S(1)–S(18)) across the inner disk, consistent with low-velocity C-type shocks (10–40 km\,s$^{-1}$) moving perpendicular to the jet axis. Their excitation diagrams are well described by a power-law temperature distribution with an exponential cutoff (see \cref{eq:modtogi}).

    \item Bright rovibrational H$_2$ lines from the $\nu=1$ and 2 levels are also detected. Their excitation deviates from thermal expectations mainly due to low densities. Nonthermal processes contribute 10–30\% of rovibrational emission, likely powered by an ADAF producing X-ray radiation.

    \item The OPRs drop to 1.5–2.5 in shocked regions. These sub-equilibrium values trace recent shocks ($\sim$0.1–1 Myr), before the OPR conversion equilibrates.

    \item Jet-driven low-velocity shocks heat the H$_2$ without disturbing the kinematics of large-scale dust lanes. In contrast, the inner $\sim$200 pc exhibit disturbed kinematics and outflow-like features, consistent with turbulent molecular gas displacement and compression spatially coincident with the edge of the circumnuclear SF ring. Double-horned H$_2$ line profiles at the shock front indicate turbulent, stratified feedback.
    
\end{enumerate}

These results demonstrate that even low-luminosity AGN can produce sustained and measurable influence on their host galaxies, not by ejecting gas through high-velocity winds, but by subtly reshaping the molecular ISM via turbulence, shocks, and thermal regulation. In M58, this “gentle” feedback thermally perturbs the inner disk and quenches SF at kiloparsec scales, while producing turbulent motions only in the center. M58 thus exemplifies the cumulative, stratified nature of low-power AGN feedback: long-lived and spatially structured, with observable imprints in gas excitation, dynamics, and SF history. 

Future studies will be essential to determine how common this scenario is in broader population of LLAGN. Systematic JWST surveys of nearby AGN-hosting galaxies, in combination with high-resolution ALMA and radio data, will be critical for quantifying the frequency, energetics, and long-term impact of weak AGN feedback in the local universe.

\begin{acknowledgements}
We thank A. Alonso-Herrero, I. Garcia-Bernete, S. Quai, M. Ceci and L. Ulivi for useful discussion. We also thank M. Dadina for providing access to his computational facilities. IEL acknowledges support from the Cassini Fellowship program at INAF-OAS. EB and GC acknowledge financial support from INAF under the Large Grant 2022 ``The metal circle: a new sharp view of the baryon cycle up to Cosmic Dawn with the latest generation IFU facilities'' and the GO grant ``A JWST/MIRI MIRACLE: Mid-IR Activity of Circumnuclear Line Emission''. EB also acknowledges financial support from ``Ricerca Fondamentale 2024'' program (mini-grant 1.05.24.07.01).  
SGB  acknowledges support from the Spanish grant PID2022-138560NB-I00, funded by MCIN/AEI/10.13039/501100011033/FEDER, EU.
The color schemes used in this work are color-blind friendly from Paul Tol's Note.

\end{acknowledgements}

\bibliographystyle{aa}
\bibliography{manuscript.bib}

\begin{appendix} 
\section{Data reduction}
\label{appendix:data_red}

Data were reduced using the JWST Science Calibration Pipeline \citep{Bushouse2025zndo..14597407B}. For NIRCam and MIRI, we used version 1.17.1 with CRDS context \verb|jwst_1321.pmap|. The NIRSpec data was processed with the JWST calibration pipeline version 1.15.1 using the CRDS context \verb|jwst_1303.pmap|. Processing parameters were tailored to each observing mode, with additional custom routines described below.

For imaging modes, we used the \verb|suppress_one_group=| \verb|False| option during ramp fitting to recover partially saturated pixels using \verb|frame0| exposures. Backgrounds were subtracted via sigma-clipped medians measured in off-source regions. NIRCam 1/f noise was suppressed using \verb|Clean_flicker_noise| from the \verb|det1dict| module, applied across all channels.

Astrometric alignment was performed using NIRCam as reference frame, given its superior positional accuracy among the JWST instruments used. The NIRSpec H$_2$ 1–0 S(1) line map was aligned to the NIRCam F212N image, yielding offsets of $\delta$RA = 0\farcs15 and $\delta$Dec = 0\farcs13 . For MIRI, the H$_2$\,S(8) line was aligned to the NIRSpec corrected frame, with offsets of $\delta$RA = 0\farcs05 and $\delta$Dec = –0\farcs08. Final cubes were resampled using the \verb|EMSM| algorithm to a spatial scale of $\sim$0\farcs1.

NIRSpec data were reduced with custom outlier rejection scripts to mitigate residual cosmic rays. We corrected for the artifacts from spatial under-sampling of the PSF, also known as "wiggles," using the algorithm by \citet{perna2023dualwiggles}. This tool was applied only within a $\sim$5-pixel radius centered on the AGN, where the effect is significant due to source brightness.

MIRI MRS data required extensive correction to address fringing and noise resampling artifacts. Fringing was removed using a dedicated 2D correction algorithm. In Channels 3 and 4, fringing correction was deactivated due to the absence of significant interference patterns. A custom correction was necessary for noise resampling, particularly in Channel 1, where the dither strategy under-sampled the PSF, producing artificial flux jumps near the central AGN. 

Using a similar idea behind the "wiggles'' algorithm, for each problematic MIRI MRS spaxel \( s(\lambda) \), a reference spectrum \( s_{\text{ref}}(\lambda) \) was extracted by averaging the spectra within a circular 5-pixel aperture centered around the spaxel. Both the spaxel and reference spectra were processed to isolate the continuum emission by masking emission lines and PAH features. The masked regions were interpolated, and a Savitzky–Golay filter of the problematic spaxel and its reference spectrum was applied to derive smooth continuum estimates \( c(\lambda) \) and \( c_{\text{ref}}(\lambda) \) respectively.

The corrected spaxel spectrum \( s_{\text{corrected}}(\lambda) \) was then calculated using the following formula:

\begin{equation}
    s_{\text{corrected}}(\lambda) = s(\lambda) \cdot \frac{c_{\text{ref}}(\lambda)}{c(\lambda)} \cdot \frac{\int c(\lambda) \, d\lambda}{\int c_{\text{ref}}(\lambda) \, d\lambda}
\end{equation}

\begin{figure*}
    \centering
    \includegraphics[width=1.0\textwidth]{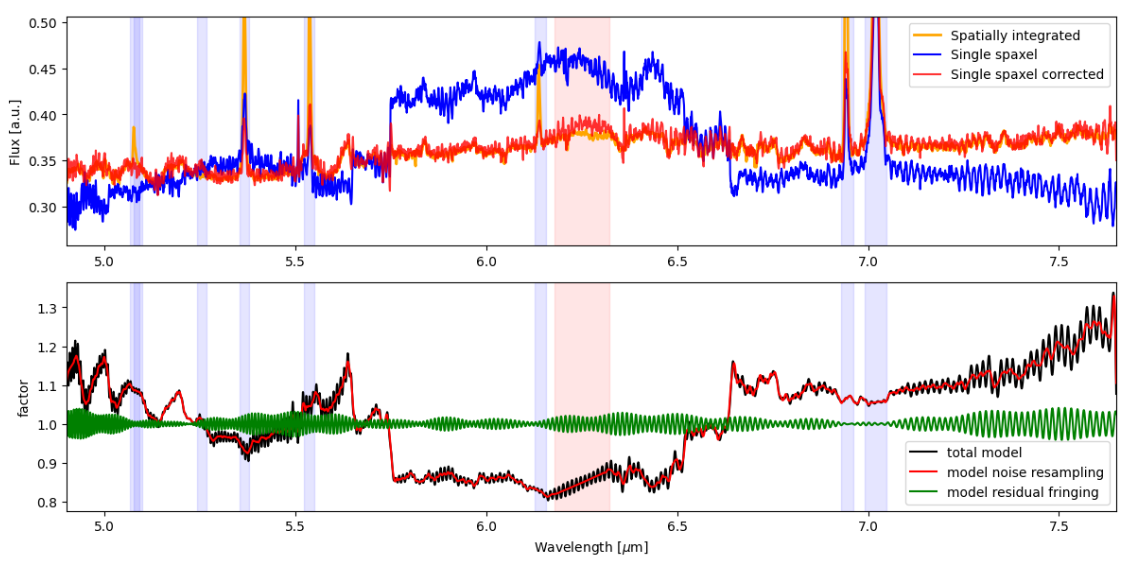}
    \caption{Example of the spaxel correction method. Top panel: Original spaxel spectrum (blue), its reference continuum (orange), and corrected spectrum (red). Bottom panel: Correction factor for each wavelength and the fringing pattern derived from the JWST pipeline. Masking lines and PAH features are shown in the shaded-blue and shaded-red areas, respectively.}
    \label{fig:correction_example}
\end{figure*}

This transformation scales the continuum of the spaxel to match that of the local reference while preserving the relative line strengths and spectral shape. A residual 1D fringe correction, derived from the JWST pipeline, was applied to mitigate remaining low-level fringing patterns. An example of this correction process is shown in Figure~\ref{fig:correction_example}, where the original and corrected spectra are compared alongside the estimated continuum shapes. 

Some MIRI spaxels exhibited artifacts that mimicked emission lines near the hydrogen recombination lines of the Humphreys series (i.e., HI 7–6). These features were discarded for two main reasons: first, the lines did not exhibit a Gaussian profile consistent with genuine emission; and second, there was no corresponding Pfund-$alpha$ line detected, which should be brighter than HI 7–6 if the signal were real.

MERLIN archival data were obtained from two observations in the C band (centered at 5,GHz, bandwidth 16,MHz) on 14 November and 4 December 2002. The total exposure times were 2.6 and 10.5 hours, respectively. The data were processed using the standard MERLIN pipeline, which includes flagging of bad visibilities, and calibration in both phase and amplitude. The two visibility datasets were combined using the \texttt{dbcon} task in \textsc{AIPS}. Imaging was performed with \textsc{DIFMAP}, using a pixel scale (cell size) of 10 mas. After several iterations of phase self-calibration followed by one iteration of amplitude self-calibration, we achieved a final image with an rms noise level of 0.19\,mJy\,beam$^{-1}$.

The VLA C-band data were taken in B configuration on 30 June 2020 (Project ID 20A-272, PI: Ogle), with a total bandwidth of 2,GHz, divided into 32 spectral windows each with 64$\times$1\,MHz channels. The raw data were calibrated using the standard VLA pipeline within \textsc{CASA} (version 5.3.1). Imaging was carried out with the \texttt{tclean} task using Briggs weighting with \texttt{robust}=0.5. The resulting synthesized beam was 0\farcs96\,$\times$\,0\farcs83, and the rms noise was 24.4\,$\mu$Jy\,beam$^{-1}$.

\section{Spectral fitting}
\label{appendix:fitting}

For the NIRSpec data, we adopted a multicomponent spaxel-by-spaxel fitting approach following \citet{marasco2020,tozzi2021} with JWST-specific updates from \citet{ceci2025,ulivi2025}. The stellar continuum and emission lines were fit simultaneously using the penalized PiXel-Fitting package \citep[pPXF;][]{cappellari2004_ppxf,cappellari2023_ppxf}. We modeled the continuum using the single stellar population templates of \citet{maraston2011} ($R$\,$\simeq$\,$20,000$ covering part of Band 2). For $\lambda$\,$>$\,$2.5~\mu$m, we used an eighth-order polynomial to approximate the redder continuum shape. Emission lines were modeled with one to three Gaussian components with linked kinematics, selected via a Kolmogorov–Smirnov (KS) test on residuals. The continuum-subtracted cubes were smoothed with a Gaussian kernel of $\sigma=1$ pixel to improve S/N and reduce outliers, at the expense of a slight loss in spatial resolution.

For MIRI MRS, we adapted the method of \citetalias{Ogle2024}\footnote{Available online at {\url{https://github.com/ie-lopez/MPF}}}, originally developed for Spitzer IRS, to the higher spatial and spectral resolution of JWST. Fits were performed with the \texttt{lmfit} Python package \citep{lmfit}, which uses Levenberg–Marquardt least-squares optimization. After masking emission lines, we fit the continuum in each spaxel with a combination of dust continuum, PAH features (modeled with templates similar to \texttt{PAHFIT}; \citealt{2007ApJ...656..770S}), and a power-law component.We then fit the residual spectra to model emission lines with one to three Gaussians, following a similar approach used for NIRSpec. We further validated the multicomponent selection with the Bayesian and Akaike Information Criteria, which yielded consistent results. The instrumental point spread function (PSF) was not explicitly accounted for during spaxel-by-spaxel fitting; however, for analyses spanning a wide wavelength range, such as excitation diagrams and the flux measurements reported in Table~\ref{table:h2}, the NIRSpec and MIRI datacubes were smoothed to match the worst PSF at the H$_2$ S(1) line (FWHM = 0\farcs67), following the wavelength-dependent PSF characterization for MRS by \citet{Law2023AJ....166...45L}. In addition, we chose apertures larger than the PSF, ensuring that PSF-related effects are minimized in the derived fluxes and line ratios. For analyses involving ratios of lines at similar wavelengths, such as the rovibrational line ratios shown in \cref{fig:grid} and discussed in \cref{subsec:H2_origin_rob}, fluxes are measured within the apertures without smoothing. This approach is more accurate because these lines are observed with the same instrument, so PSF mismatches are negligible and only contribute to slightly larger uncertainties.

We verified that both methods yield consistent results for lines covered by both instruments, such as the H$_2$~S(8) transition. Given the superior sensitivity and resolution of NIRSpec, we adopt its measurement for this line. Finally, all integrated spectra were extracted directly from the original cubes, without applying "wiggles" correction, to avoid interpolation artifacts. When integrating over apertures larger than the PSF, the wiggles average out and become negligible. These integrated spectra were modeled using the same spaxel-by-spaxel fitting procedure.

\section{H$_2$ detected lines}
\label{appendix:lines}
\begin{figure}
    \centering
    \includegraphics[width=1.0\linewidth]{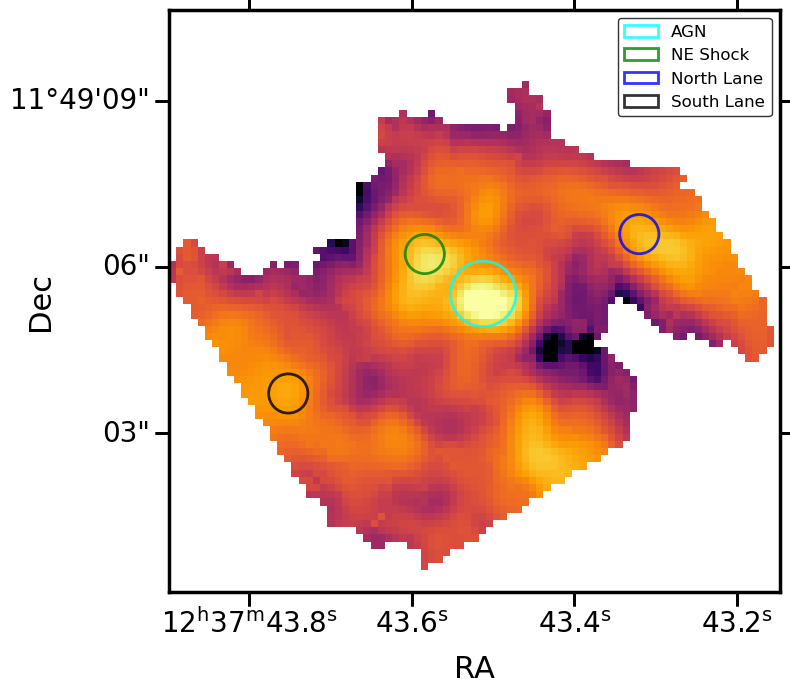}
    \caption{H$_2$ S(5) line map with selected extraction regions. The central aperture (cyan) is slightly enlarged to match the PSF, while the surrounding apertures sample the NE shock lobe and dust lanes.}
    \label{fig:regions}
\end{figure}

Table~\ref{table:h2} lists the H$_2$ emission lines detected in each of the selected regions indicated in \cref{fig:regions} . All reported fluxes are above the 3$\sigma$ significance threshold. The central AGN region is defined by a 0.5\farcs\ radius aperture centered on the MERLIN radio core (RA = 189.4313, Dec = 11.8182), encompassing the compact AGN and matched to the MIRI MRS PSF. Four additional circular apertures of 0.3\farcs\ radius sample distinct morphological and kinematic features:
NE shock (RA = 189.4316, Dec = 11.8184),
North Lane (RA = 189.4305, Dec = 11.8185), and
South Lane (RA = 189.4323, Dec = 11.8177). 

\begin{table*}
\label{table:h2}
\caption{Molecular hydrogen lines detected.}
\centering
\begin{tabular}{lcccccccc}
\hline \hline
Transition & Notation & 
\begin{tabular}{c} 
$\lambda$ \\ 
\end{tabular} & 
\begin{tabular}{c} 
$\frac{E_u}{k}$
\end{tabular} & 
\begin{tabular}{c} 
$A_\mathrm{tot}$\\ 
\end{tabular} &Nucleus&Northeast Shock&North Lane&South Lane\\
\noalign{\smallskip}
(1) & (2) & (3) & (4) & (5) & (6) & (7) & (8) & (9) \\
\hline
\hline
\noalign{\vskip 1.5mm}\multicolumn{5}{l}{$\nu=0 \longrightarrow 0$}\\
\noalign{\vskip 0.5mm}  
$J=3 \rightarrow 1$ &   S(1)  & 17.035 & 1015  & 47.6   & 11.85 (0.51) & 4.60 (0.07) & 7.96 (0.03) & 10.46 (0.02)\\
$J=4 \rightarrow 2$ &   S(2)  & 12.279 & 1681  & 276    & 18.63 (0.85) & 6.37 (0.11) & 7.64 (0.03) & 9.06 (0.02)\\
$J=5 \rightarrow 3$ &   S(3)  & 9.665  & 2503  & 984    & 51.83 (1.57) & 19.90 (1.01) & 17.83 (0.38) & 20.41 (0.20)\\
$J=6 \rightarrow 4$ &   S(4)  & 8.025  & 3475  & 2643   & 35.33 (0.51) & 8.14 (0.08) & 6.18 (0.67) & 7.04 (0.40)\\
$J=7 \rightarrow 5$ &   S(5)  & 6.910  & 4586  & 5879   & 90.00 (0.79) & 23.00 (0.16) & 9.91 (0.96) & 15.33 (0.02)\\
$J=8 \rightarrow 6$ &   S(6)  & 6.109  & 5830  & 11400  & 23.10 (0.41) & 5.46 (0.14) & 2.43 (0.02) & 3.42 (0.08)\\
$J=9 \rightarrow 7$ &   S(7)  & 5.511  & 7197  & 20000  & 68.00 (1.38) & 13.15 (0.38) & 4.51 (0.12) & 7.32 (0.14) \\
$J=10 \rightarrow 8$ &  S(8)  & 5.053  & 8677  & 32360  & 18.24 (1.00) & 4.31 (0.37) & 1.51 (0.15) & 1.59 (0.11) \\
$J=11 \rightarrow 9$ &  S(9)  & 4.695  & 10261 & 49000  & 37.44 (0.92) & 7.62 (0.28) & 2.65 (0.15) & 3.01 (0.17) \\
$J=12 \rightarrow 10$ & S(10) & 4.410  & 11940 & 70290  & 8.22 (0.62) & 1.58 (0.49) & 0.48 (0.05) & 0.53 (0.05) \\
$J=13 \rightarrow 11$ & S(11) & 4.181  & 13703 & 96330  & 21.29 (0.92) & 3.88 (0.66) & 1.14 (0.03) & 1.31 (0.04) \\
$J=15 \rightarrow 13$ & S(13) & 3.846  & 17444 & 161700 & 11.79 (0.62) & 2.16 (0.11) & 0.51 (0.02) & 0.52 (0.03) \\
$J=16 \rightarrow 14$ & S(14) & 3.724  & 19403 & 199900 & 3.16 (0.89) & 0.67 (0.10) & 0.08 (0.01) & 0.12 (0.02) \\
$J=17 \rightarrow 15$ & S(15) & 3.626  & 21411 & 240700 & 5.59 (0.63) & 1.53 (0.14) & 0.27 (0.02) & 0.23 (0.02) \\
$J=18 \rightarrow 16$ & S(16) & 3.548  & 23459 & 283100 & < 2.71 & 0.59 (0.11) & 0.08 (0.02) & 0.07 (0.03) \\
$J=19 \rightarrow 17$ & S(17) & 3.486  & 25539 & 325900 & 2.71 (0.60) & 0.95 (0.13) & < 0.08 & 0.08 (0.02) \\
$J=20 \rightarrow 18$ & S(18) & 3.439  & 27643 & 368000 & 3.03 (0.55) & 1.10 (0.14) & 0.10 (0.02) & < 0.08 \\
\noalign{\vskip 1.5mm}\multicolumn{5}{l}{$\nu=1 \longrightarrow 0$}\\
\noalign{\vskip 0.5mm}  
$J=2 \rightarrow 0$ & 1-0 S(0) & 2.223 & 6471 & 84720  & 12.47 (2.54) & 2.01 (0.17) & 0.66 (0.06) & 1.57 (0.19) \\
$J=3 \rightarrow 1$ & 1-0 S(1) & 2.122 & 6951 & 83800  & 34.97 (1.53) & 7.03 (0.24) & 2.56 (0.07) & 2.89 (0.07) \\
$J=4 \rightarrow 2$ & 1-0 S(2) & 2.034 & 7584 & 82260  & 18.54 (2.52) & 3.11 (0.22) & 1.03 (0.06) & 1.02 (0.07) \\
$J=5 \rightarrow 3$ & 1-0 S(3) & 1.958 & 8365 & 80110  & 74.80 (28.11) & 8.15 (0.30) & 2.75 (0.09) & 2.96 (0.33) \\
$J=6 \rightarrow 4$ & 1-0 S(4) & 1.892 & 9286 & 77630  & 73.54 (15.23) & 2.95 (0.28) & 0.92 (0.07) & 3.12 (0.27) \\
$J=7 \rightarrow 5$ & 1-0 S(5) & 1.836 & 10341 & 75330 & 44.27 (3.70) & 6.22 (0.34) & 2.15 (0.09) & 2.42 (0.29) \\
$J=8 \rightarrow 6$ & 1-0 S(6) & 1.788 & 11521 & 73970 & < 22.76 & < 1.64 & < 0.46 & 0.86 (0.36) \\
$J=9 \rightarrow 7$ & 1-0 S(7) & 1.748 & 12817 & 74430 & 37.67 (6.73) & 3.18 (0.40) & 0.99 (0.12) & 1.92 (0.45) \\
$J=0 \rightarrow 2$ & 1-0 O(2) & 2.627 & 5987 & 85320  & 12.55 (1.60) & 1.58 (0.43) & 0.53 (0.04) & 0.47 (0.04) \\
$J=1 \rightarrow 3$ & 1-0 O(3) & 2.803 & 6149 & 85160  & 19.98 (0.98) & 4.92 (3.46) & 1.55 (0.06) & 1.31 (0.07) \\
$J=2 \rightarrow 4$ & 1-0 O(4) & 3.004 & 6471 & 84720  & 6.67 (0.61) & 2.01 (0.19) & 0.46 (0.03) & 0.63 (0.04) \\
$J=3 \rightarrow 5$ & 1-0 O(5) & 3.235 & 6951 & 83800  & 12.67 (0.60) & 2.47 (0.06) & 0.93 (0.02) & 1.04 (0.03) \\
$J=4 \rightarrow 6$ & 1-0 O(6) & 3.501 & 7584 & 82260  & 3.78 (0.56) & 0.68 (0.05) & 0.20 (0.02) & 0.23 (0.02) \\
$J=5 \rightarrow 7$ & 1-0 O(7) & 3.807 & 8365 & 80110  & 6.05 (0.57) & 1.44 (0.11) & 0.39 (0.02) & 0.35 (0.02) \\
$J=7 \rightarrow 9$ & 1-0 O(9) & 4.575 & 10341& 75330  & 2.11 (0.72) & 0.49 (0.23) & 0.05 (0.06) & 0.29 (0.09) \\
$J=7 \rightarrow 7$ & 1-0 Q(7) & 2.500 & 10341& 75330  & 11.24 (1.18) & 2.40 (0.16) & 0.86 (0.11) & 1.12 (0.28) \\
$J=8 \rightarrow 8$ & 1-0 Q(8) & 2.528 & 11521& 73970  & < 14.85 & < 0.72 & 0.18 (0.07) & < 1.47 \\
$J=9 \rightarrow 9$ & 1-0 Q(9) & 2.560 & 12817& 74430  & 9.34 (3.59) & 0.91 (0.26) & 0.49 (0.09) & 0.46 (0.09) \\
$J=11 \rightarrow 11$ & 1-0 Q(11) & 2.635 &15722&84630 & 5.19 (0.74) & < 2.93 & 0.16 (0.03) & 0.27 (0.04) \\
\noalign{\vskip 1.5mm}\multicolumn{5}{l}{$\nu=1 \longrightarrow 1$}\\
\noalign{\vskip 0.5mm}  
$J=11 \rightarrow 9$ & 1-1 S(9)   & 4.954 & 15722 & 84630 & 5.88 (0.93) & 0.60 (0.15) & 0.25 (0.10) & 0.42 (0.10) \\
$J=12 \rightarrow 10$ & 1-1 S(10) & 4.656 & 17311 & 95980 & < 7.32 & 0.23 (0.18) & < 0.45 & 0.13 (0.09) \\
$J=13 \rightarrow 11$ & 1-1 S(11) & 4.417 & 18979 & 112300& 3.48 (0.68) & 0.44 (0.11) & < 0.26 & < 0.34 \\
\noalign{\vskip 1.5mm}\multicolumn{5}{l}{$\nu=2 \longrightarrow 1$}\\
\noalign{\vskip 0.5mm}  
$J=3 \rightarrow 1$ & 2-1 S(1) & 2.248 & 12550 & 162300 & 6.88 (3.72) & 1.18 (0.16) & 0.33 (0.05) & < 1.00 \\
$J=5 \rightarrow 3$ & 2-1 S(3) & 2.073 & 13890 & 157200 & 7.85 (2.59) & 1.52 (0.20) & 0.32 (0.05) & 0.46 (0.06) \\
$J=2 \rightarrow 2$ & 2-1 Q(2) & 2.559 & 12095 & 163000 & 5.06 (2.65) & 0.95 (0.23) & 0.13 (0.08) & < 0.42 \\
$J=3 \rightarrow 3$ & 2-1 Q(3) & 2.570 & 12550 & 162300 & 4.15 (2.38) & < 0.91 & 0.27 (0.19) & 0.20 (0.10) \\
$J=1 \rightarrow 3$ & 2-1 O(3) & 2.974 & 11789 & 163100 & < 5.31 & 0.44 (0.17) & 0.12 (0.04) & < 0.35 \\
\hline
\end{tabular}
\tablefoot{(1) Transition level, (2) Notation, (3) Vacuum wavelength in microns, (4) Excitation energy in K, (5) Einstein A coefficient in units of $10^{-11}\,\mathrm{s}^{-1}$, (6-9) Fluxes and their 1-$\sigma$ errors for different regions in unit of $10^{-16}\,\mathrm{erg}\,\mathrm{cm}^{-2}\,\mathrm{s}^{-1}$. Upper limits are calculated at 5-$\sigma$. H$_2$ transition data from \citet{Roueff2019A&A...630A..58R}.}
\end{table*}

\end{appendix}
\end{document}